\RequirePackage[hyphens]{url}

\documentclass[superscriptaddress,amsmath,amssymb,aps,twocolumn,pre]{revtex4-2}

\usepackage{graphicx}
\usepackage{dcolumn}
\usepackage{bm}

\usepackage[c]{esvect}
\usepackage{upgreek}
\usepackage{placeins}
\usepackage{lipsum}

\newcommand{\vy}[1]{\protect\vv{\textbf{\text{#1}}}} 

\let\Phi\varPhi
\let\Theta\varTheta

\begin{document}

\title{A generalized vector-field framework for mobility}

\author{Erjian Liu}\email{erjian@ifisc.uib-csic.es}
\affiliation{Instituto de F\'{\i}sica Interdisciplinar y Sistemas Complejos IFISC (CSIC-UIB), 07122 Palma de Mallorca, Spain} \affiliation{School of Systems Science, Beijing Jiaotong University, Beijing 100044, China}

\author{Mattia Mazzoli}\affiliation{ISI Foundation, 10126 Turin, Italy}

\author{Xiao-Yong Yan}\affiliation{School of Systems Science, Beijing Jiaotong University, Beijing 100044, China}

\author{Jos\'e J. Ramasco}\email{jramasco@ifisc.uib-csic.es}\affiliation{Instituto de F\'{\i}sica Interdisciplinar y Sistemas Complejos IFISC (CSIC-UIB), 07122 Palma de Mallorca, Spain}

\begin{abstract}
Trip flow between areas is a fundamental metric for human mobility research. Given its identification with travel demand and its relevance for transportation and urban planning, many models have been developed for its estimation. These models focus on flow intensity, disregarding the information provided by the local mobility orientation. A field-theoretic approach can overcome this issue and handling both intensity and direction at once. Here we propose a general vector-field representation starting from individuals' trajectories valid for any type of mobility. By introducing four models of spatial exploration, we show how individuals' elections determine the mesoscopic properties of the mobility field. Distance optimization in long displacements and random-like local exploration are necessary to reproduce empirical field features observed in Chinese logistic data and in New York City Foursquare check-ins. Our framework is an essential tool to capture hidden symmetries in mesoscopic urban mobility, it establishes a benchmark to test the validity of mobility models and opens the doors to the use of field theory in a wide spectrum of applications. 
\end{abstract}

\maketitle


Characterizing mobility patterns between locations of people and goods is not only a long-standing research topic in disciplines such as geography and spatial economics \cite{wilson1970,bergstrand1985,karemera2000gravity,roy2003spatial,rouwendal2004,barthelemy2011spatial,carra2016,barbosa2018human}, but it has also critical practical applications for urban planning \cite{batty2013,barthelemy2016,li2017simple,bassolas2019hierarchical}, building and expansion of transportation infrastructure   \cite{zipf1946p,stouffer1940intervening,kitamura2000micro,ortuzar2011}, population distribution studies and city sprawl \cite{deville2014dynamic,louail2014mobile,louail2015uncovering,verbavatz2020growth,pappalardo2021evaluation}, urban socioeconomic development \cite{xu2018human,barbosa2021uncovering,mimar2022connecting}, location-based services \cite{scellato2011exploiting} and infectious diseases spreading  \cite{viboud2006,balcan2009,balcan2011,tizzoni2014,jia2020population,mazzoli2021interplay,aguilar2022impact}. 

Early research works mainly used data from transportation surveys and census to analyze people travel and activity patterns \cite{boyce2015,barbosa2018human}. The collection of such mobility data can be expensive and time-consuming, and it is hard to acquire a sufficient sample size \cite{levinson1995activity,axhausen2002observing}. With the development of information and communication technologies (ICTs), the availability of large-scale high-resolution mobility data, such as  call detailed records, GPS-located taxi data and online social network data, has notably increased. This enabled researchers to quantitatively study various mobility contexts and put forward new models to analyze the underlying mechanism of mobility  \cite{gonzalez2008understanding,bagrow2012,noulas2012,lenormand2014,hawelka2014,lenormand2015, blondel2015survey,barbosa2018human}.

In terms of models, mobility can be studied at two different levels: individual trajectories and aggregated flows between areas. Individual mobility models usually focused on characterizing the behavior of individuals in the process of selection of destinations, adding a certain degree of stochasticity to account for people heterogeneity and free will (see \cite{brockmann2006scaling,gonzalez2008understanding,song2010modelling,alessandretti2020scales} for some examples and \cite{barbosa2018human} for a recent review). At the aggregated level, the earliest models fall into two families: the gravity  \cite{carey1867,zipf1946p} and the intervening opportunity model \cite{stouffer1940intervening,ruiter1967}, which has later evolved into the so-called radiation model \cite{simini2012universal}. These models and their updated versions predict flows between locations, describing travel distance distribution and defining locations attractiveness \cite{simini2012universal,yan2014universal,lenormand2016systematic,liu2020universal,simini2021deep,yan2019destination}. However, they are not designed to capture the local flow orientation, which is a spatial information that  plays a significant role on describing mesoscopic mobility patterns \cite{bongiorno2021vector,shida2020universal}. 

Such combination of mobility flow intensity and orientation can be studied using a field theoretical framework. The idea of using fields and potentials for studying mobility emerged in the context of the gravity model, where these concepts appear in a natural way \cite{stewart1947, mukherji1975mobility}. The lack of data prevented further advances on this direction, until a recent work \cite{mazzoli2019field} proved that a vector field framework could be used to characterize trips between home and work (commuting) in a number of cities in the world. Not only that, this framework was able to solve a controversy almost 80 years old on which of the two families of models performs best to describe commuting. The gravity model produces results that matches empirical commuting mobility patterns both in intensity and orientation of the flows. A field representation has lately been used in a machine learning context, where knowing the potential can significantly improve the performance of the model to predict traffic flows \cite{wang2022traffic}. Later, some works translated the field approach to lower scales of mobility (single flows) \cite{aoki2021urban,shida2020universal, shida2022approximation}, pedestrian route selection \cite{bongiorno2021vector} or the mobility associated to the celebration of special events \cite{yang2022vector}. Nevertheless, it is not yet clear how the definition of mesoscopic mobility fields can be extended to any type of mobility starting from individual trajectories and what are the features that may permeate from the microscopic mobility information to the mesoscopic scale. 

\begin{figure*}
\centering
\includegraphics[width=16cm]{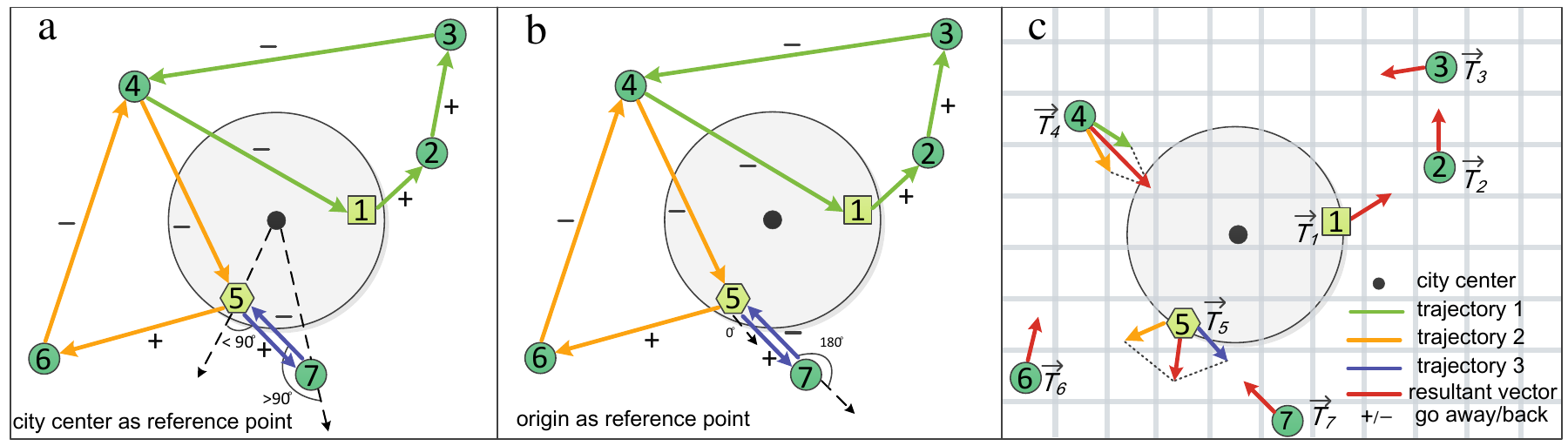}
\caption{\textbf{Definition of trajectories orientation and resulting vector field.} The large gray circle stands for an idealized city (with the black point as the city center), the green numbered circles are the stops sequence of each trajectory (1-2-3-4-1, 5-6-4-5 and 5-7-5), while the square in 1 and the hexagon in 5 are the trajectories' origins. {\bf a} City center as RP. Vectors of each color connect consecutive stops of a trajectory, for example, the trajectory 1-2-3-4-1. When the vector connecting $X$ to $Y$ forms an acute angle with the position vector $\protect\vy{X}$ starting in the city center, we mark the vector $\protect\vy{XY}$  as positive, meaning that the agent is moving away from the city center, like the vector $\protect\vy{57}$. Vice versa, we mark the vector as negative when the agent moves towards the city center, e.g. the vector $\protect\vy{75}$. {\bf b}  Trajectory-origin as RP. Identically to what we defined for the city center but we use the origins of trajectories $1$ or $5$, respectively, as RP for the position vectors. {\bf c}  Sketch of the method to build the vector field. The space is divided in a grid, vectors departing from stops in a cell $i$ are normalized and summed vectorially to produce $\protect\vy{T}_{i}$. Then we define the mobility field $\protect\vy{W}_i$ dividing the vector $\protect\vy{T}_i$ by the number of trips departing from $i$.}
\label{fig1}
\end{figure*}

In this work, we introduce a definition for the vectorial framework of mobility encoded in individual trajectories  valid for all mobility types. We investigate
how individuals' choices determine the features of the mobility vector field. We find that empirical trajectories extracted from logistic motivated trips in Chinese cities and Foursquare check-ins in New York City lead, following our definition, to well-behaved vector-fields, which satisfy the divergence (Gauss's) theorem and have no curl. We propose three individual mobility models and analyze what is the minimal set of ingredients needed to find fields similar to the empirical ones. Our results show that a distribution of stops decaying with the distance to the city, length optimization for long displacements and random-like local exploration are fundamental to reproduce empirical mobility fields. This work extends thus the mathematical amenability of mobility data and offers a new approach for individual mobility patterns analysis.

\section*{Results}

\subsection*{From trips to vectors}

A sketch of the method to define a vector field is displayed in Fig. \ref{fig1}, where we show an idealized ``circular'' city, its central point (black circle) and three trajectories: 1-2-3-4-1, 5-6-4-5 and 5-7-5. Most of our trajectories are closed, although this is not necessary for the method to work. The steps to define the vector field are as follows:
\begin{enumerate}
\item We define vectors between consecutive stops. This can be seen, for example, in Fig. \ref{fig1}{\bf a} and \ref{fig1}{\bf b}, where the vectors $\vy{12}$, $\vy{23}$, $\vy{34}$ and $\vy{41}$ can be extracted from the first trajectory. A vector pointing from one stop $X$ to the next $Y$ is called $\vy{XY}$ and is located on $X$. 
\item The vectors $\vy{XY}$ are normalized to obtain the unit vectors $\vy{xy}$.
\item The space is divided in grid cells of equal area and all unit trip vectors $\vy{xy}$ within each cell $i$ are vectorially summed to define the resulting vector $\vy{T}_i$, which informs on the average mobility direction in $i$ (see Fig. \ref{fig1}{\bf c}). 
\item  $\vy{T}_i$ is normalized by the total number of trips leaving cell $i$ to obtain the mesoscopic mobility vector field $\vy{W}_i$. This process is analogous to defining the gravitational or electrical fields dividing the force by the mass or charge, respectively. 
\end{enumerate}
The vectors $\vy{W}$ constitute thus the mobility field.

\subsection*{From trip vectors to trajectory orientation}

In order to characterize how individuals explore space we need to determine whether trips in each area head towards or away from a given reference point (RP). We identify two options for RP: the first one is the city geographical center (Fig. \ref{fig1}{\bf a}), in this case the RP is absolute and equal for all the trajectories; the second option is to establish the origin of each trajectory as RP (Fig. \ref{fig1}{\bf b}). This latter option implies that the RP is different for every trajectory. As we will see below, the trajectory-origin RP shows some useful features and most of the results here are, therefore, displayed using such RP unless otherwise stated. 

With respect to the chosen RP, we can allocate a sign to each displacement vector $\vy{XY}$ (or $\vy{xy}$) of any trajectory. Remember that the vector $\vy{XY}$ sits on $X$, and that every stop $X$ can be described by a position vector $\vy{X}$ from the RP to $X$. To understand whether the displacement $\vy{XY}$ occurs toward or away from the RP, we compute the angle $\theta$ in the range $(-180,180]$ between the above two vectors (see Fig. \ref{fig1}{\bf a} and \ref{fig1}{\bf b}), where $|\theta| = 0$ means moving straight away from the RP and $|\theta| = 180$ means moving strictly toward the RP. We assign the vector $\vy{XY}$ a positive sign ($+$) if $|\theta|< 90$, and a negative sign otherwise ($-$). By convention, vectors $\vy{XY}$ are positive if the stop $X$ coincides with the RP. Some examples from Fig. \ref{fig1} are the vectors $\vy{34}$ or $\vy{64}$ that are both negative pointing to the RP in both representations, or the vector $\vy{23}$ that is positive. Vectors exiting from the origins such as $\vy{57}$, $\vy{12}$ and $\vy{56}$ are by convention positive. 

We characterize for every trajectory $t$ whether trips are on average toward ($-$) or away ($+$) from the RP by summing the signs of all the displacement vectors. Dividing by the total number of vectors, we obtain the average orientation $H_t$ laying in the range between $-1$ and $1$, where $H_t=1$ means that all the displacements are positive (i.e., away from the RP) and vice-versa for $H_t=-1$. Note that distance or duration of trips are not taken into account. 
Finally, $H_t=0$ implies a full balanced trajectory. $H_t$ is a microscopic observable that encodes individual behavior features of mobility.

A priori, there is no reason to assume that there should be more displacements in one orientation than in the other. This means that overall there should be as many trajectories with positive or negative $H_t$. We count as $N_+$ the number of trajectories with $H_t >0$, $N_-$ the number of those with $H_t <0$ and $N_0$ as the number of balanced trajectories. We finally introduce the unbalance ratio $\rho$ as
\begin{equation}
\rho = \frac{N_+}{N_-}.
\end{equation}
The existence and orientation of the mesoscopic field and the orientation of trajectories are mathematically interlinked. Since every displacement generates an unit trip vector contributing to the overall mobility field, a situation with $\rho$ clearly deviating from one could generate a majority direction for displacements and, consequently, an average field (see Appendix D for a demonstration that $\rho < 1$ implies the existence of a field).

\begin{figure}
\centering
\includegraphics[width=8cm]{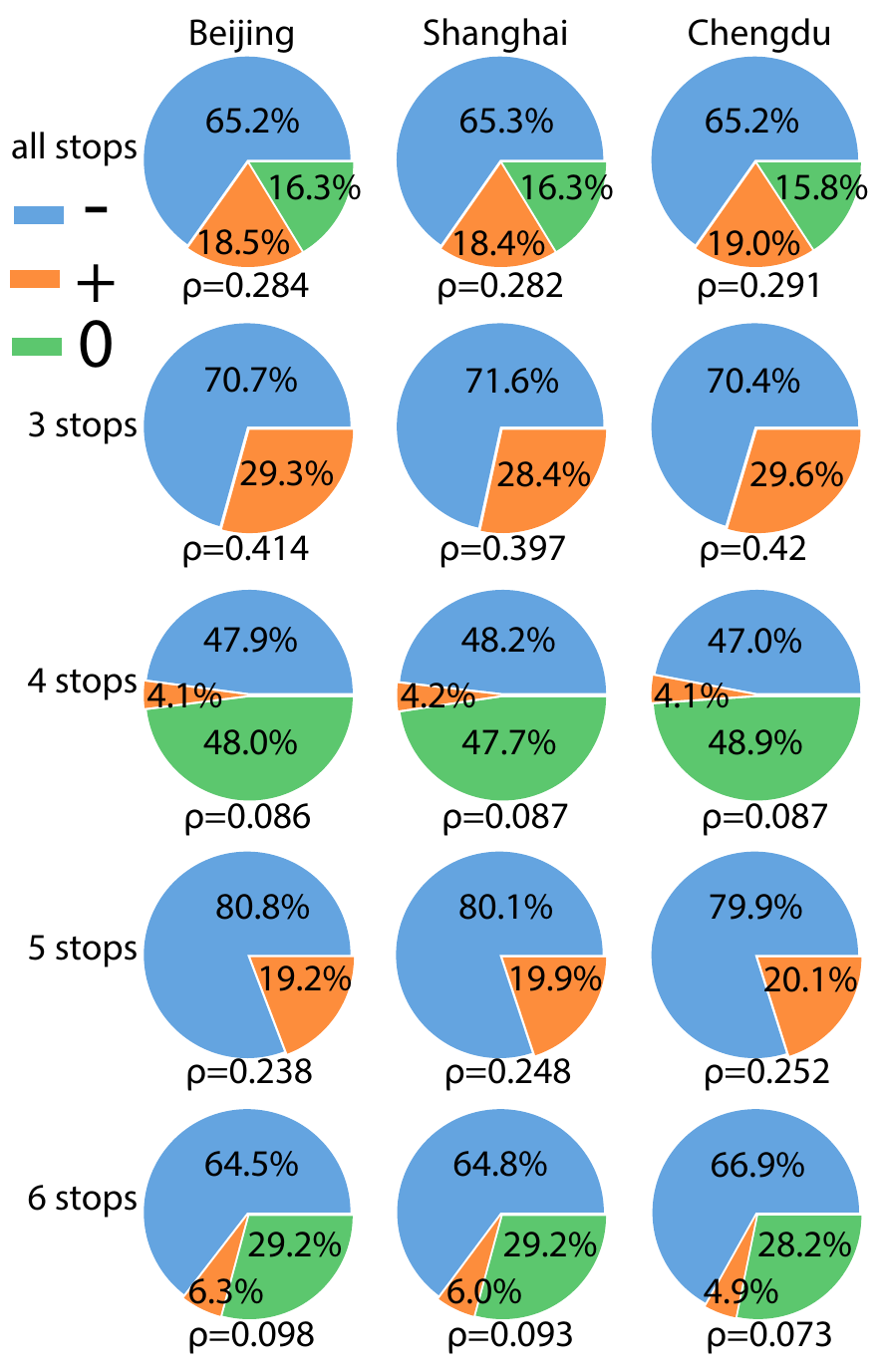}
\caption{{\bf Unbalance of empirical trajectories.} Fraction of positive (orange), $H_t > 0$, negative (blue), $H_t < 0$ and balanced (green), $H_t = 0$ trajectories with $3$, $4$, $5$, $6$ stops and for all trajectories of trucks departing from Beijing, Shanghai and Chengdu, using the trajectory origin as RP. $\rho$ stands for the ratio between the number of positive and negative trajectories.
}
\label{fig2}
\end{figure}

\subsection*{Empirical trajectories orientation}

Knowing that an unbalance in the trajectories may lead to the presence of a mobility field, it is important to test whether empirical trajectories are  balanced or not. We consider two datasets: the first one, D1, refers to logistic trucks trajectories departing from the $21$ largest Chinese cities, while the second one, D2, refers to Foursquare check-ins of individuals in New York City (NYC) (see Methods and Appendix A for a detailed description of the two datasets).
We show the empirical results for $\rho$ in Beijing, Shanghai and Chengdu of D1 using trajectory origins as the RP in Fig. \ref{fig2}. The pie charts show the fraction of positive, negative and balanced trajectories with a fixed number of stops and also the overall results. Trajectories with negative orientation (i.e., $N_-$)  are dominant, hence $\rho < 1$ in all cases. Results are robust for trajectories of any number of stops and for all trajectories in all the $21$ cities analyzed. This reflects that goods delivery trucks tend to reach the furthest location first and then gradually approach the origin (so that $H_t < 0$). 
Note that negative trajectories dominate independently from the length of trajectories and there is high similarity in the $\rho$ values across cities, overall and for trajectories with a certain number of stops. This result does not hold if the RP is the city center (see Appendix C and Figs. S5-S7). It seems that the analysis performed with the origins of trajectories as RP is able to absorb the details of the cities in terms of shape, streets and communication axes (e.g. highways), hence leaving only individual mobility behavior. This has two consequences: firstly, we can neglect the urban shape when modeling individuals' mobility behavior in terms of $\rho$; secondly, we can tune the model on an arbitrary city and perform out-of-sample accurate predictions. The results are robust for further cities (see below in Understanding the origin of the trajectory unbalance section and Appendix E Figs. S10-S12). 

Since the definitions of vector signs and trajectory orientations are general, we can apply it to any type of mobility. As a comparative, we perform the same analysis on D2, the check-in records of Foursquare, and find a similar pattern (see Fig. S10).  Note that D2 encodes a different type of mobility compared to D1. The values of $\rho$ for D2 with individuals' check-ins are different from the ones obtained from D1 with freight data both for trajectories of a certain stops number only or for all trajectories.

\subsection*{Understanding the origin of the trajectory unbalance}

\begin{figure}
 \centering
\includegraphics[width=8cm]{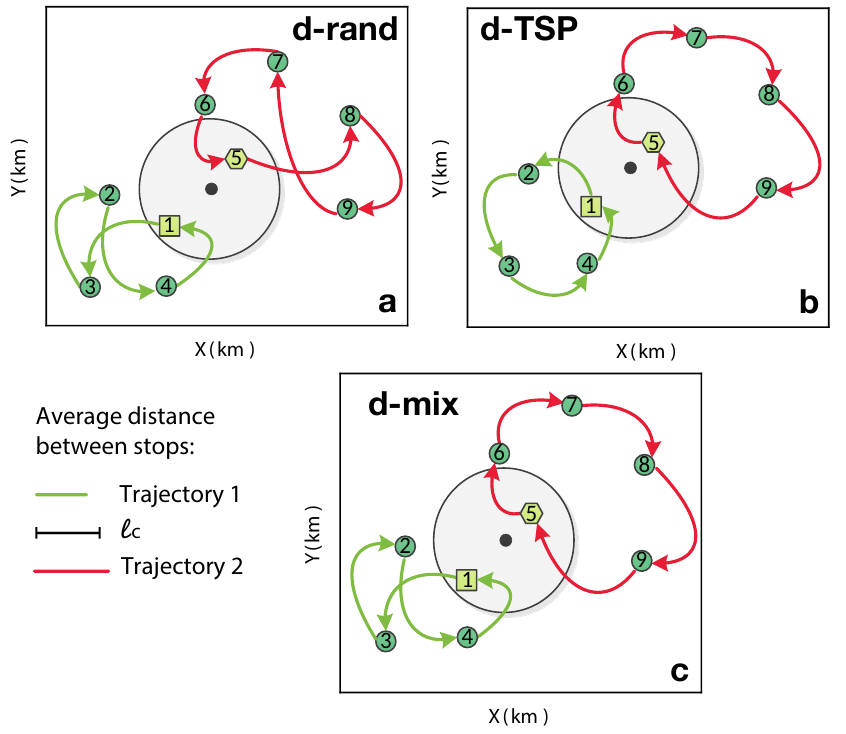}
\caption{{\bf Schematic description of the models.}  The large gray circle represents the city area (the black circle is the city center), the numbered green circle represent stops while square and hexagon are the origins of two trajectories (red and green). These trajectories have different average trips distance: the green one is below $\ell_c$, the red trajectory is above $\ell_c$. \textbf{a} d-rand, \textbf{b} d-TSP, \textbf{c} d-mix.}
\label{fig3}
\end{figure}

In order to understand the mechanisms leading to the empirically observed unbalance between the signs of the trajectories, we introduce four models in an increasing order of complexity, each with added mechanisms over the previous one to characterize the needed ingredients (see Methods for details). The simplest configuration includes a circular city of radius $R$. A generic trajectory is composed of a sequence of stops $\{\vy{x}_0, \vy{x}_1,... \vy{x}_{s-1}\}$, with the origin $\vy{x}_0$ located randomly inside the circular city.

For the first model, called Rand, the other stops locations are selected completely at random in space. We have confirmed that the model trajectories tend to be balance in the thermodynamic limit (large $L$) and that it does not generate a field  (see Fig. S15). This was a sanity check before advancing to more elaborated models and we disregard this model from now on. The next three models are more relevant and are developed making simple behavioral assumptions about spatial navigation, a sketch with their description can be seen in Fig. \ref{fig3}. 

The d-rand model follows the same logic but is informed with a spatial distribution of stops $D(r)$ decaying with the distance to the city center. This mimics a random-like exploration but with constraints on the spatial distribution of stops. The next model, d-TSP, allows travelers to reorder the stops to minimize the total distance traveled (with a Traveling Salesman Problem (TSP) optimization algorithm). Finally, d-mix interpolates between the two previous models and the distance optimization only occurs if the average distance between trajectory stops goes over a threshold $\ell_c$. Trajectories distances are not optimized otherwise. More details on the precise definition of the models are offered in the Methods section below. 

The models are informed by the empirical statistics from Beijing (see Methods for further details on the models construction). We tune the d-mix model parameter $\ell_c$ by minimizing the mean absolute error on $\rho$ in Beijing using all trajectories and the trajectory origin as RP. The best results are obtained for $\ell_c \approx 1.72 \, km$, which is a reasonable threshold for route optimization (see Appendix J and Fig. S20).

\begin{figure*}
\centering
\includegraphics[width=14cm]{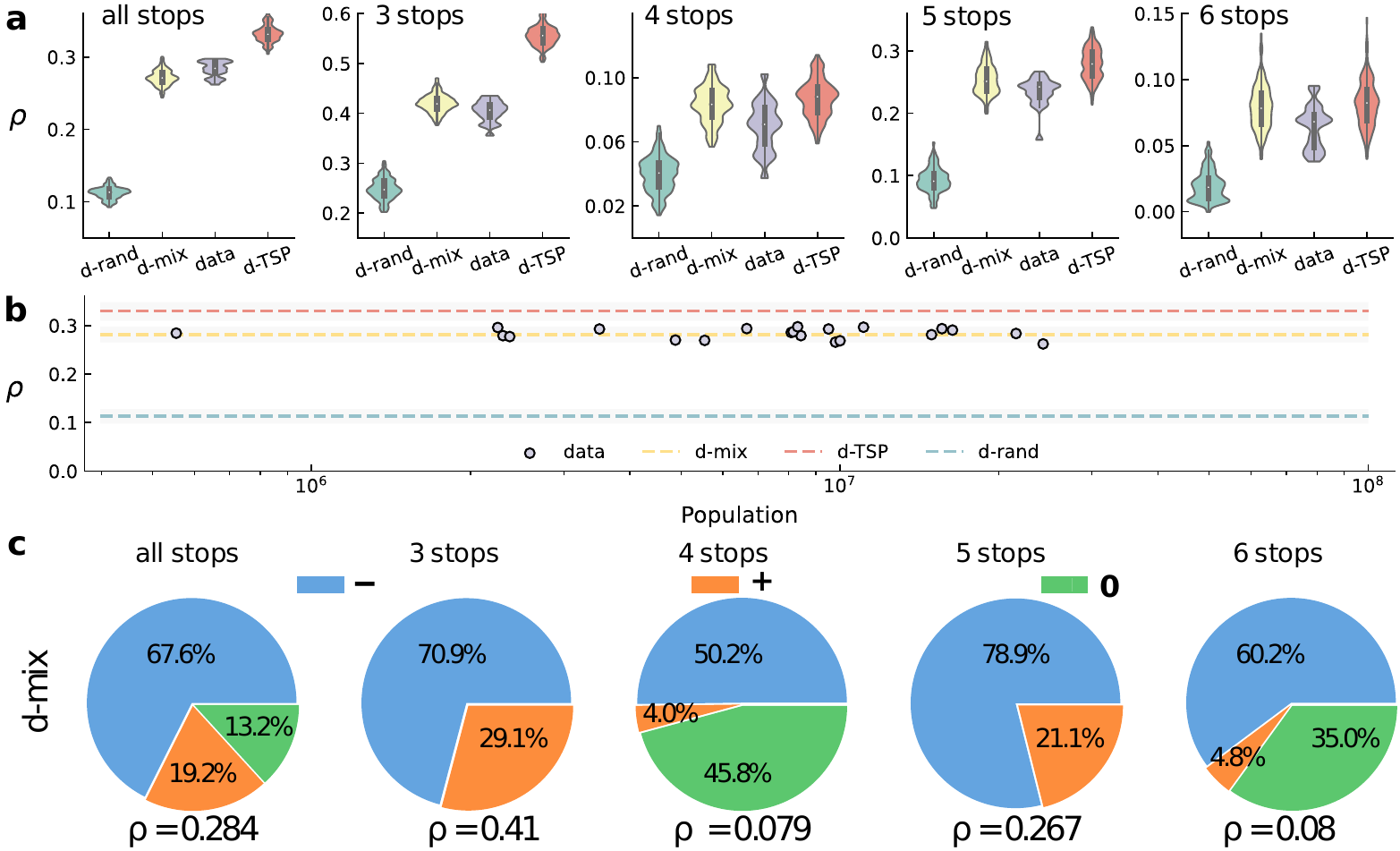}
\caption{{\bf Models and trajectory unbalance. } {\bf a}  Violin plots of the model-predicted and real values of $\rho$ for trajectories with different number stops. The violin plots depict the values of $\rho$ from d-rand model (in green), d-mix model (in yellow), the data in D1 (in purple) and d-TSP model (in red) for trajectories with 3, 4, 5, 6 stops and for all trajectories. {\bf b}  Values of $\rho$ for the different cities as a function of their population. The horizontal dashed lines correspond to the median values of $\rho$ for all the modeled trajectories, shaded areas indicate the confidence interval between $5\%$ and $95\%$, color codes as above. Purple dots represent the values of $\rho$ for all empirical trajectories in each city. \textbf{c} Pie charts for the fraction of positive, negative and balanced trajectories generated with the fitted d-mix model. These sequence of charts should be compared with the empirical values observed in different cities in Fig. \ref{fig2}. For these simulations, we used $R = 20 \,km$ and $L = 400\, km$  and with $100\ 000$ trajectories. } 
\label{fig4}
\end{figure*}

Violin plots in Fig. \ref{fig4}\textbf{a} show the resulting distribution of $\rho$ for the three stochastic models and the empirical distribution from the 21 cities in D1. We also show the distributions for trajectories with a fixed amount of stops only. The first question to highlight is that all the models produce trajectories that are consistently negative and hence holding $\rho < 1$. A second aspect is that d-rand generates the lowest values of $\rho$ among all models. This is natural since the location of consecutive stops is random, without ordering them to reduce the distance traveled, and hence, the probability of crossing to the other side of the city (a negative sign for the trip) is high. Many negative trips contribute to an overall negative sign for the trajectories and a lower value of $\rho$. Models with stops reordering may reduce the number of displacements from one side to the other of the city, and the trajectory signs are less often negative (higher values of $\rho$). In contrast and following the same argument in the opposite way, d-TSP produces trajectories with the highest value of $\rho$. $\rho$ for d-mix, on the other hand, lies between these two extremes as do also the empirical values of $\rho$. When the analysis is restricted to trajectories with a certain number of stops, the stochastic fluctuations are larger since the number of trajectories decreases. However, d-mix fits best to the empirical values of $\rho$. 

In  Fig. \ref{fig4}\textbf{a} we show the empirical $\rho$ for trajectories from all the 21 cities in D1 together. The cities contributing to this violin plot are, nevertheless, very heterogeneous in population. This is why in Fig. \ref{fig4}\textbf{b} we depict the empirical values $\rho$ as a function of population and see that there is no noticeable dependence. Moreover, all the empirical values fluctuate within the $95\%$ interval of the value of $\rho$ obtained by the d-mix model (fitting $\ell_c$ only in Beijing). Finally, we can analyze the trajectories generated by the fitted d-mix model by their number of stops. In Fig.\ref{fig4}\textbf{c}, we see that, in general, the agreement for trajectories of a fixed number of stops aligns well with the empirical results from the three cities in Fig. \ref{fig2}. All these results have been confirmed using different values of the modeled city size $R$ and the space considered $L$ (see Methods for details on the models' parameters and Appendix G (Fig. S15) for the robustness check).

\subsection*{The d-mix model mobility field}

\begin{figure}
\centering
\includegraphics[width=8cm]{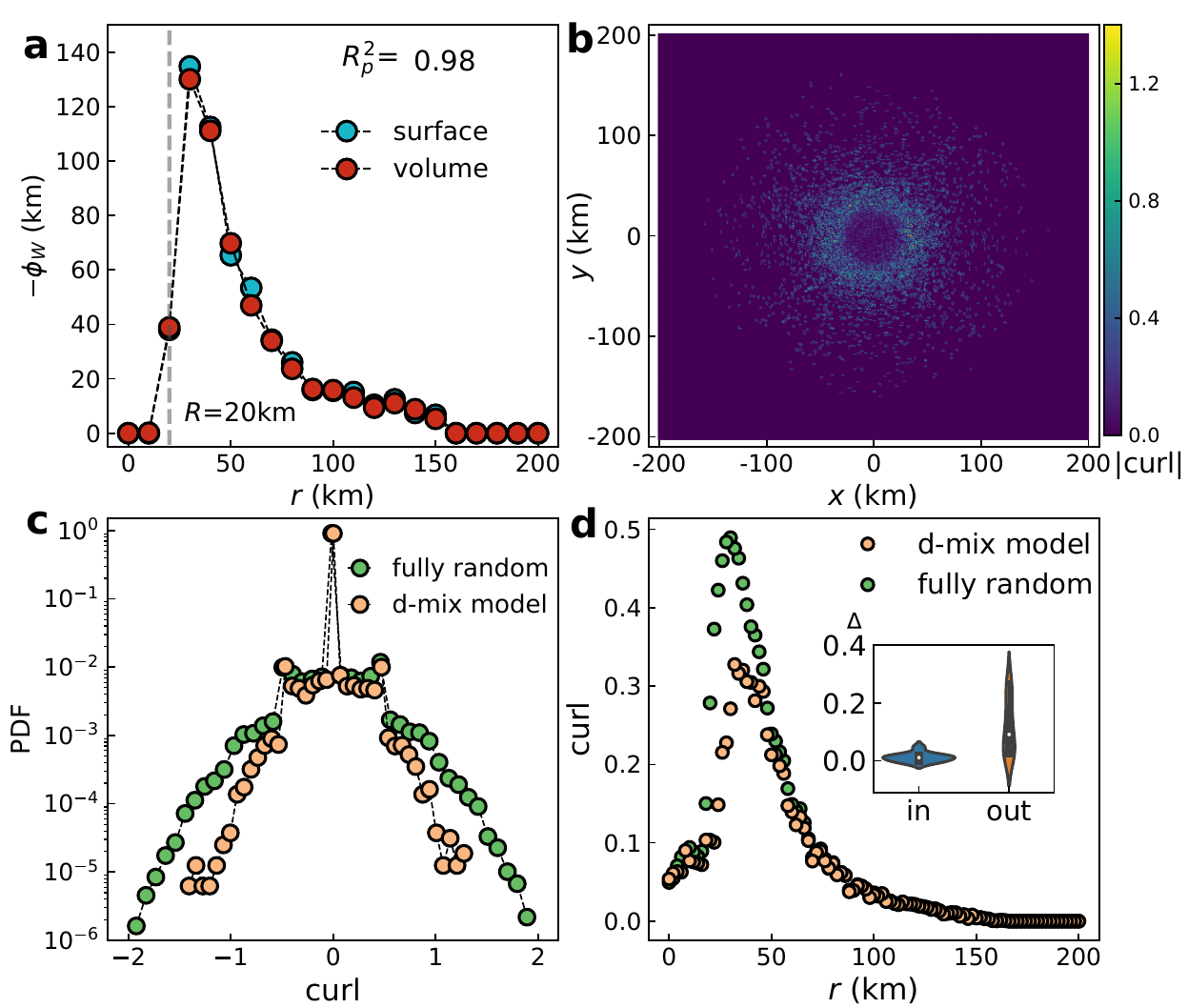}
\caption{{\bf Properties of the d-mix mobility field.} The mobility field is obtained from the d-mix model with $\ell_c = 1.72 \, km$ with a circular city of radius $R = 20 \, km$ in a space limited by a box of side $L = 400 \, km$ and with $100\, 000$ trajectories. {\bf a} Comparison between the flux measured as the surface integral in blue, and as the volume integral of the divergence of the mobility field (see Methods for the  calculation) in red, in both cases as a function of the distance to the city center $r$. The coefficient of determination ${R^2}_p$ is obtained as the square of the Pearson correlation coefficient of both curves. {\bf b} Module of the curl of the field, the colors represent $|\nabla \vy{W}|$ for each cell in ${km^{-1}}$. {\bf c}  Comparison between the curl of the field generated with the d-mix model and that of the fully random model obtained by randomly reassigning directions to $\vy{W}$ in each cell. {\bf d} Average absolute value of the curl as a function of the distance to the city center for the d-mix and the fully random model.}
\label{fig5}
\end{figure}

Since the models produce unbalanced trajectories, i.e. $rho < 1$, they also generate a net mobility field (see Appendix D for a mathematical proof). Next we study the properties of the field $\vy{W}$ generated by the fitted d-mix model (see Methods for the details of the modeling setting). We consider circular contours of radius $r$ from the city center and analyze the flux of $\vy{W}$ as a function of $r$. The flux is calculated in both as a surface integral over the circular contour  and as the volume integral of the divergence of the mobility field, $\nabla \, \vy{W}$ (see Methods for the formal flux calculation). According to Gauss' Divergence Theorem, if the field is well behaved the two ways of calculating the flux should yield the same result. This is confirmed in Fig. \ref{fig5}\textbf{a}, where the two calculations of the flux $\Phi_W$ as a function of the distance are in agreement with a coefficient of determination $R^2_P = 0.98$. Note that this does not apply to the city area where the estimated fluxes are close to zero. The fact that the Gauss' Theorem is fulfilled is important because it is related to the existence of a source for the field.

A second relevant feature to explore is the field curl. This is connected to the possibility of defining a potential for the field. Fig. \ref{fig5}\textbf{b} displays the module of the curl, which lies in the z-axis perpendicular to the plot. One can distinguish the area of the city in the internal circle. There are some non-zero curl areas, with the highest values concentrated close to the city border. Actually, these values are small when compared with a null model (see Fig. \ref{fig5}\textbf{c}). In this null model, the direction of the d-mix vectors in each cell is randomly reoriented. We call this the ``fully random'' model and it is intended to assess the level of curl induced only by noise. The overall distribution of curl modules of the field generated by the d-mix has lower variance than the null model one. Furthermore, in Fig. \ref{fig5}\textbf{d}, we compare the average module of the absolute value of the curl enclosed by a circle of radius $r$ from the city center. We see that both models coincide inside the city $r \le R = 20 \, km$. However, beyond the city area the d-mix model has curl values systematically below the null model ones. This guarantees the possibility to define a potential out of the city for the d-mix mobility field.

\subsection*{Empirical mobility fields}

\begin{figure}
\centering
\includegraphics[width=8cm]{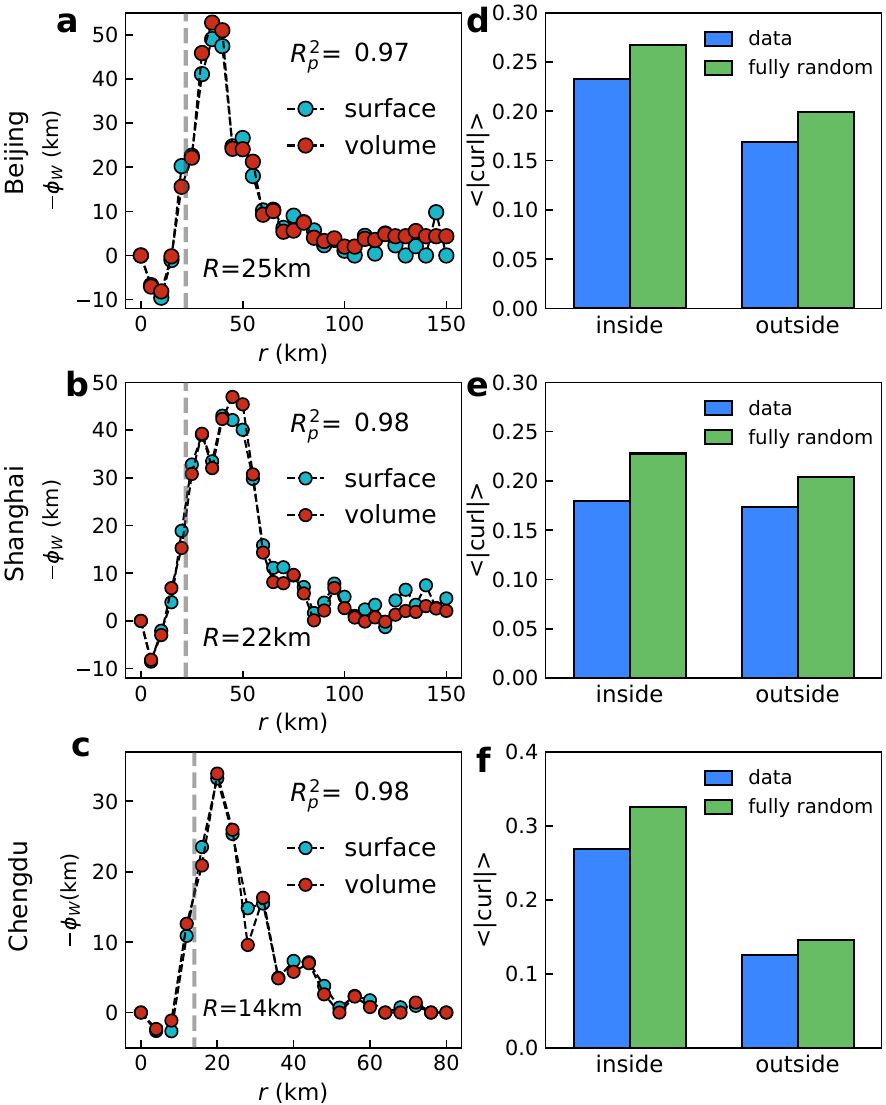}
\caption{{\bf Empirical vector fields.} In {\bf a-c}, fluxes of the empirical field $\vy{w}$ as a function of the distance to the city center $r$. In blue the surface integral flux, in red the volume integral flux, both as a function of $r$. $R_p^2$ is the coefficient of determination between the curves of the flux and the volume integral of the field divergence as a function of the distance . {\bf d-f} Average module of the curl, comparing the fully-random model and the empirical data. The separation between in and out of corresponds to the distance at which $\Phi_W = \Phi_{\rm max}/2$, $r_c$, as described in the text. The analysis in the outer side ends at $r = 50 \, km$. After this distance, trip vectors in Beijing and Chengdu become sparse and the statistics is non-representative.}
\label{fig6}
\end{figure} 

In Fig. \ref{fig6}\textbf{a-c}, we see that the empirical fields generated in the three cities used as example fulfill the Gauss' Theorem as well. Results for further cities are included in Fig. S16.  The flux profiles as a function of $r$ show some common features: a first $r$ interval with negative $\Phi_W$ values (vector field pointing mainly outwards). This area is the core of the city, where the logistic trucks drop goods while most of the origins of the trajectories lie further in the peri-urban area. Despite its negative character, the existence of a net flux is relevant because the agreement between both integrals show that the Divergence Theorem is fulfilled in all the space. Further, the flux becomes positive (vector field pointing on average inwards) and it reaches the half height value at $r_c = 25 \, km$ for Beijing, $r_c = 22 \, km$ for Shanghai and $r_c = 14\, km$ for Chengdu. Since for the d-mix model the city size approximately coincides with the point at which $\Phi_W(R) \approx \Phi_{\rm max}/2$, we will take $r_c$ as arbitrary radii of the three cities logistic cores. The fluxes peak further away ($\sim 35\, km$ in Beijing, $\sim 50\, km$ in Shanghai and $\sim 20\, km$ in Chengdu) and eventually $\Phi_W$ decays as $r$ increases. These general features compare well with those of the d-mix model (Fig. \ref{fig6}\textbf{a}), except for the inside city fluxes, which are not well captured by the model. 

We display in Fig. \ref{fig6}\textbf{d-f} the average absolute value of the curl enclosed and excluded by a circle of radius $r_c$, i.e. internal and external respectively, to the cities. We find that the empirical curl is systematically lower than the fully-random counterpart in all cases. This implies that empirical potentials can be defined in all the space. Similar results are attained with other cities of D1 (see Fig. S17) and with the Foursquare check-in data in New York City (see  Fig. S18).   

\begin{figure}
\centering
\includegraphics[width=8cm]{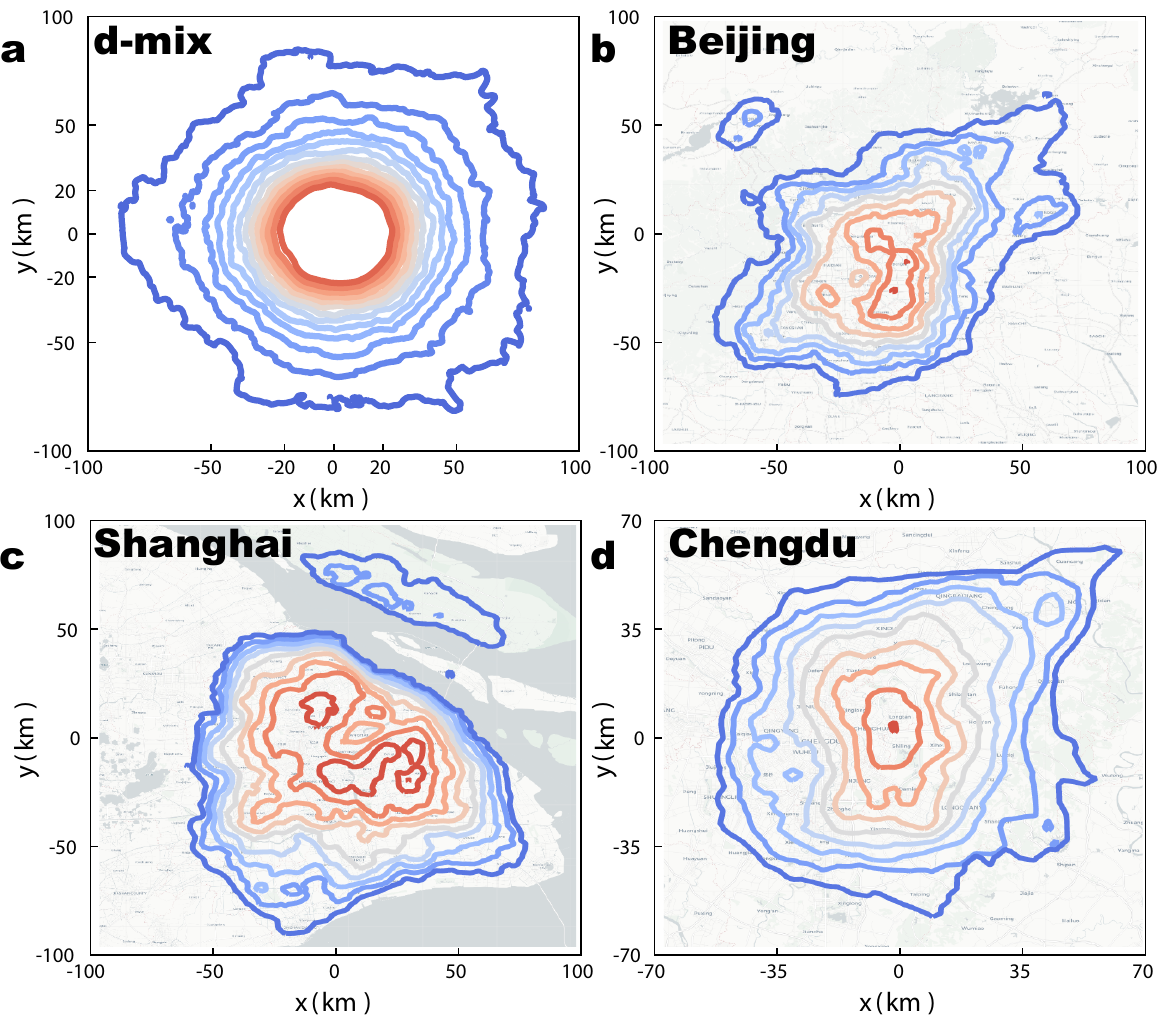}
\caption{{\bf Equipotential contours.} {\bf a} Contours for the d-mix model with $100\, 000$ trajectories in a circular city of radius $R = 20 \, km$ and in a box of side $L = 400 \, km$. Empirical equipotential contours for  {\bf b}
 Beijing, {\bf c} Shanghai and {\bf d} Chengdu.}
\label{fig7}
\end{figure}

\subsection*{Potentials}

Knowing that we can define a potential for the d-mix model (out of the city) and also for the empirical data everywhere, we plot next the equipotential curves on the maps (Fig. \ref{fig7}). The potential of the d-mix model shows a circular symmetry. This is due to the circular city shape introduced (Fig. \ref{fig7}\textbf{a}) and the isotropic assumption. The contours of the empirical urban areas are dependent on the city shape (Fig. \ref{fig7}\textbf{b-d}), adapting to geographical constraints as in the case of Shanghai with the sea and islands. It is also interesting how the potential highlights the presence of satellite cities as occurs for Beijing and Shanghai.  The potential contours plotted extend some tens of kilometers outside the cities, eventually becoming fuzzy. The field continues beyond that point, but given the lack of statistics it is hard to extract meaningful potential contours.

\begin{figure*}
\centering
\includegraphics[width=14cm]{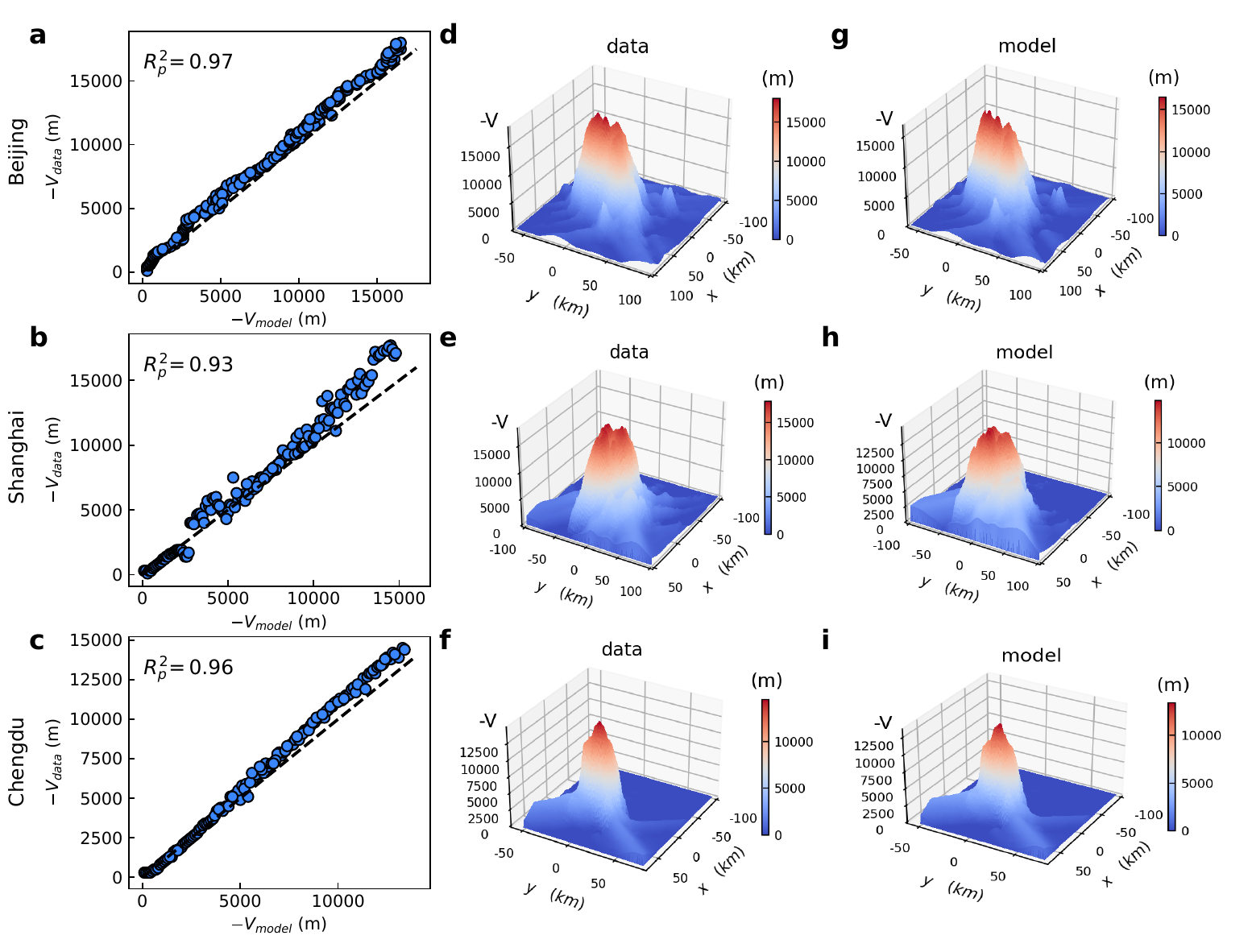}
\caption{{\bf  Hybrid d-mix model predictions of empirical potentials. } {\bf a-c} Correlation plots between the hybrid d-mix model and the empirical potentials, yielding ${R_p}^2=0.97$ for Beijing, ${R_p}^2=0.93$ for Shanghai and ${R_p}^2=0.96$ for Chengdu. {\bf d-f} Empirical potential in the space and {\bf g-i} the hybrid d-mix model predictions. Both models and empirical flows show the polycentric nature of Beijing and Shanghai, while Chengdu is more monocentric.}
\label{fig8}
\end{figure*}

\subsection*{Hybrid d-mix model}

We saw how the unbalance ratio of trajectories orientation $\rho$ does not depend on the urban shape when using trajectory origins as RP. This is critical since it allowed us to study the origin of the mobility fields with simplified models. However, the spatial shape of the fields and potentials are inherently connected to the real city configuration. To explore further the validity of the model assumptions, we need to make another step and introduce a hybrid d-mix model. We consider the empirical trajectories one by one, e.g., $\{ \vy{x}_0, \vy{x}_1, ...,  \vy{x}_{s-1} \}$, keeping $\vy{x}_0$ as origin, but randomizing the order of the other stops. We then input these trajectories to the d-mix model, reordering them according to the model rules. The resulting trajectories are not necessarily equal to the empirical original ones, although if the model is doing a good work they should be similar. Indeed, we have a coincidence of $79 \%$ in Beijing, $71\%$ in Shanghai and $79\%$ for Chengdu. The question is thus whether this over $20\%$ mismatch has mesoscopic effects on the field or not. 

The potential estimated from the hybrid d-mix  model and the empirical one are compared for the three cities in Fig. \ref{fig8}\textbf{a-c}. We find a good agreement with $R_P^2$ above $0.93$ for all the three cities. 3D profiles of the potential are displayed in Figs. \ref{fig8}\textbf{d-f} for the empirical fields and in Figs. \ref{fig8}\textbf{g-i} for the field generated by the  hybrid d-mix model. One can clearly appreciate the similarity between modeled and empirical potentials of the same city. This implies that the hybrid d-mix model is capturing well the mechanisms behind the empirical mobility fields and its utility may go beyond its use as explanatory tool for individual behavior as above. The major deviations occur in the largest potential values, close to the maxima and the city centers where the d-mix potential is undefined and for which the model has not been fitted. 

\section*{Discussion}
In this work we have introduced a new way to define a mobility field starting from individual trajectories. This is a major generalization with respect to previous works based on Origin-Destination commuting matrices, and it allows us to study a wide range of mobility data. Besides the conceptual leap, with this new framework we have studied mobility data from two different sources: logistic routes of trucks around and across the 21 largest Chinese cities and Foursquare check-ins in NYC. In all cases, we have found a well-behaved field fulfilling the Gauss Divergence Theorem and with a curl value that it is in general smaller than the one expected by a fully random model. This implies that it is possible to define a potential almost anywhere in metropolitan areas and, consequently, to search for a source for the mobility field.

Starting from individual behavioral assumptions of spatial exploration, we have advanced in the conceptual framework by analyzing the basic ingredients needed to generate mesoscopic mobility fields with features matching those of the empirical ones. We have introduced a metric, the unbalance ratio $\rho$, to characterize the fraction of displacements in trajectories that move mostly towards or away from a reference point RP (city center or origin of the trajectory). The unbalance among these directions ($\rho$ far from one) implies a net displacement direction and induces a mobility field. This metric allows us to quantify the strength of the factors leading to the formation of the field. We have then introduced a set of minimal models with growing complexity to explore what information is fundamental to generate the field. All the models are based on trajectories, and so the basic components are: an origin $\vy{x}_0$ and a sequence of stops $\{\vy{x}_1,... \vy{x}_{s-1}\}$, with $s$ representing the total number of locations visited in the trajectory. The simplest model, a random selection of the origin within the city and random location for the stops, is not able to generate a field. We have then added an ingredient: stops randomly extracted following a decaying distribution of the distance to the city center, and built the d-rand model. This model generates trajectories with an unbalance ratio less than the unit and, consequently, a mobility field. However, it is not able to reproduce the empirical values of $\rho$ (its $\rho$ is smaller). To approach realistic values, one must include the fact that individuals aim at optimizing their trajectories, reordering the sequence of stops $\{\vy{x}_0, \vy{x}_1,... \vy{x}_{s-1}\}$ to reduce the total distance traveled. Mimicking this process, we added to d-rand a Traveling Salesman Problem solver to reduce total distance, and built the d-TSP model. This model generates a field with $\rho$ closer to the empirical one than the produced by d-rand model, but still higher. The assumption of rational optimization of all trajectories does not hold for short distance trips. For this reason we introduced the d-mix model, which optimizes stops only if the average displacement between them is larger than a given threshold $\ell_c$. d-mix model interpolates between both behaviors, and can be tuned on $\ell_c$ to generate trajectories with a $\rho$ consistent with the empirical ones. d-mix model is not only able to reproduce $\rho$, but most of its field features are realistic as well. The Gauss Divergence Theorem is fulfilled and the curl of the field is smaller than in a fully random field. The model also generates a potential for the field, which, however is based on isotropic assumptions and hence, may differ from the empirical ones due to urban shape and natural constraints. Moreover, the model assumes an isotropic circular city setting. This limitation is overcome by the introduction of a hybrid d-mix model informed with real but randomized trajectories stops from the data. By letting the d-mix rules apply to reorder them, this hybrid model is able to reproduce the spatial shape of the empirical fields and potentials.

This work has an eminent conceptual side, advancing on the understanding of how the field theory can be applied to the mesoscopic scales of human mobility. Field theory is a fundamental tool in physics with a well equipped set of mathematical results developed for its use, which we hope can be translated to mobility studies in the near future.   
Additionally, urban poly-centrism and predominant patterns among mobility centers have been recently the focus of many studies due to their association to life quality indexes, city livability \cite{bassolas2019hierarchical}, walkability, sustainability, services accessibility and epidemic outbreak susceptibility \cite{aguilar2022impact}.
The potential provides a clear representation of the structure of a city at a mesoscopic scale.
It captures the spatial organization and connectivity patterns of mobility centers, offering insights into the distribution of activities, resources and flows. This information can help urban planners to take more informed decisions.  
 
\section*{Methods}

\subsection*{Mobility data and processing}

The empirical results of this work are based on two datasets: the first is a truck travel records (D1) \cite{yang2022identifying, yang2022identifyingE}, and the second is the check-in records of Foursquare (D2) \cite{bao2012location}. 

The D1 dataset includes data from over 20 Chinese cities, which can be downloaded from the National Road Freight Supervision and Service Platform (https://www.gghypt.net/). 
This platform is used to record the real-time geographic locations of all heavy trucks in China and monitor the potential traffic threats. The dataset contains more than 2.7 million  travel records, spanning from May 18, 2018 to May 31, 2018. The attributes of one travel record include truck ID, timestamp, longitude, latitude and speed (see Appendix A for more details about the dataset used in this study).

The D2 dataset  is from  New York city \cite{bao2012location}.
Foursquare is a location-based social network on which users share their coordinates when check-in in (see appendix A for more details about the dataset).
This dataset contains  42035 individuals, in which 23520 users have trips among different analysis zones (here the space is divided according 2010 census areas, see https://www.census.gov/geo/maps
data/maps/block/2010/). 

\subsection*{Definition of the RP in D1 and D2}

In both D1 and D2, we can define the RP as the city center or the trajectories origins. For the latter case, in D1 this is identified as the truck most commonly visited location with the longest stay times, since this is likely to be the logistic center of operations. In D2, we assign individuals' origins as the location with the largest number of check-ins. In both D1 and D2, a single trajectory is the defined as the sequence of stops occurring between the first and the next stop in the origin. The statistical description of the D1 trajectories in several cities of D1 are provided in Fig. S3, and the same  for D2 in NYC in the Fig. S4. 

\subsection*{Models}

The basic ingredients of our models will be inspired by the structure and statistics of the empirical trajectories. Fig. S3 shows the complementary cumulative distributions of three variables associated to trajectories starting in a circle of radius $20\, km$ centered at Beijing, Shanghai and Chengdu as paradigmatic examples. The first distribution, $D(r)$, refers to distance of the trajectory stops to the city centers (Fig. S3\textbf{a-c}). The city centers are the barycenters of the areas considered (see Table S2). As the figure shows, the location of the stops can be relatively far away from the city center, with the distribution falling slowly to the thousands of kilometers. We will adopt in the models a probability $D(r)$ of finding a stop at a certain distance $r$ of the city center that on very first approximation will fall as $D(r) \sim r^{-\alpha}$. The next distribution, Fig. S3\textbf{d-f}, refers to the number of stops per trajectory $N_s(s)$. The minimum number of stops is $2$, because we are counting at least origin and final destination. The range of values is relatively limited, up to $40$, and presents a decay that we will approach in the models by $N_s(s) \sim s^{-\beta}$. Finally, Fig. S3\textbf{g-i} shows the distribution of the number of trajectories starting from the same origin $N_t(t)$. Truck fleets may have an operation center from which several vehicles leave, or simply the same truck appears in several trajectories starting always from the same origin. The distribution is wide, reaching more than one thousand trajectories and we will approximate it by $N_t(t) \sim t^{-\gamma}$. The distributions for the Foursquare check-in in New York City can be found  in Fig. S4.  

If we consider as in Fig. \ref{fig1}\textbf{a} circular city of radius $R$ inside of a space limited by a square of side $L$, we can build a first null model by selecting the location of the origin of a trajectory at random in the internal circle $\vy{x}_o$. We call this model the Rand model. The number of trajectories departing from $\vy{x}_o$ is then obtained as a stochastic extraction of $N_t(t)$, while for each of them the number of stops $s$ can be extracted from $N_s(s)$. The location of the $s-1$ consequent stops is randomly chosen within the bounding square. We take a setting with a circular city of radius $R = 20 \, km$ centered in a squared area of side $L = 400 \, km$. This is to be considered the general setting on which we run all our models. The model is run to generate $100\,000$ trajectories and with them produce a field $\vy{W}$ following the recipe of Fig. \ref{fig1}. In each perimeter position, we observe that the flux of the field as a function of $r$ fluctuates around zero inside the city circle $R$ and it only gets negative as $r$ reaches close to the bounding box $L$ (see Fig. S13). Such negative net flux is only a finite-size effect as can be seen by increasing the box size $L$ and also by using periodic boundary conditions in the bounding box instead of open ones (Fig. S14).

\subsubsection*{The d-rand}

There are, therefore, missing ingredients in this basic model to be able to generate a stable mobility field. The first mechanism that we are going to consider is a spatial distribution of stops falling with the distance to the city center $D(r)$ as the one observed in Fig.  S3\textbf{a} ($D(r) = 1.2 \, r^{-2.2}$ for $r \ge 1 \, km$, if $r < 1 \, km $ it is uniform ($D(r) = \frac{1}{R^2}$). This model will be called d-rand (Fig.  \ref{fig3}\textbf{a}) and it consists in randomly extracting, as before, a location in the circle containing the city for $\vy{x}_0$, the number of trajectories starting at $\vy{x}_0$ from $N_t(t)$ and the number of stops per trajectory $s$ from $N_s(s)$. Then, for each trajectory we choose at random with $D(r)$ the radius of the location of the $s-1$ stops besides the origin, the directions from the center in which every stop lies is also randomly selected. A trajectory is thus formed by the origin and all the other stops $\{ \vy{x}_0,\vy{x}_1, ... , \vy{x}_{s-1} \}$. As we will see, this model is able to produce unbalanced trajectories and a field. 
 
d-rand has, however, a major caveat: consecutive stops can be at opposite sides of the city and it is unrealistic to have a driver passing back and forth through the city center without grouping nearby stops to reduce the total distance traveled and the fuel consumed.

\subsubsection*{The d-TSP}
The next model to consider, called d-TSP (Fig. \ref{fig3}\textbf{b}), corresponds to the effect of a manager looking at the sequence $\{ \vy{x}_0, \vy{x}_1, ... \vy{x}_{s-1}\}$ obtained as in the d-rand and reordering the sequence of stops from $1$ to $s-1$ to minimize the total trajectory distance. This process can be mapped into the well known traveling salesman problem (TSP), in which a salesman needs to visit a set of locations, each location is visited once and only once, and finally must return to the starting position \cite{bellman1962dynamic}. We employ in the d-TSP an heuristic algorithm (genetic algorithm \cite{weile1997genetic}) developed to approximate the solution of the traveling salesman problem.

\subsubsection*{The d-mix}

Finally, we introduce a model that interpolates between d-rand and d-TSP. We will call this model d-mix and the rules are as illustrated in Fig. \ref{fig3}\textbf{c}. The TSP reordering of  stops is only allowed if  the average travel distance of one trajectory between stops is larger than a threshold $\ell_c$. The idea behind d-mix is that the driver will not invest the effort of optimizing the trajectory if the distance between consecutive stops is very short. The limit of d-mix model for small $\ell_c$ corresponds thus to d-TSP model, while for large $ \ell_c$ it becomes d-rand model. 

\subsection*{Numerical calculation of the flux}

The definition of the flux is
\begin{equation}
\Phi_w^s=\oint d\ell \, \vy{n}\,\vy{W},
\end{equation}
where the integral is over the surface (perimeter $S$) enclosing the area of interest, $d\ell$ is the element of surface, $\vy{n}$ the unit vector normal to the surface and $\vy{W}$ the vector field.
From a numerical perspective, the integrals are calculated as
\begin{equation}
\Phi_w^s=\sum_{i\in S} \, d\ell \,  \vy{n}_i\,  \vy{W}_i,
\end{equation}
where the index $i$ runs over all the cells intersecting the perimeter $S$, $\vy{n}_i$ is the unit
vector normal to the surface in $i$ and $d\ell$ is approximated by the total perimeter of $S$ divided by the number of intersecting cells.

The definition of the integral of the divergence is
\begin{equation}
\Phi_w^v=\int dV \,\nabla\vy{W},
\end{equation}
where the integral is now of volume (surface in 2D of the enclosed area).
From a numerical perspective, the integrals are calculated as
\begin{equation}
\Phi_w^v=\sum_{i\in V} \bigl(\frac{{W}_{x_{(\alpha+1,\beta)}}-{W}_{x_{(\alpha,\beta)}}}{\Delta x}+\frac{{W}_{x_{(\alpha,\beta+1)}}-{W}_{x_{(\alpha,\beta)}}}{\Delta y}\bigl) dV,
\end{equation}
where the index $i$ runs over the cells in the
volume $V$ and $dV$ is approximated by the area of the unit cell.
${W}_{x}$ and ${W}_{y}$ are the $x$ and $y$ components of the vector $\vy{W}$. The indices $(\alpha, \beta)$ refer to the position of cell $i$ in the grid, in such a way that, for instance, $(\alpha \pm 1, \beta)$ are the positions of the adjacent cells to $i$ in the x-direction. $\Delta x$ and $\Delta y$  are side sizes of the cells in the $x-$ and $y-$ directions. 

\subsection*{Numerical calculation of the  curl}

The curl of $\vy{W}$ in the cell $i$, whose indices in the $x-$ and $y-$ directions are ($\alpha$, $\beta$),  as above, is determined as:
\begin{equation}
\nabla\times\vy{W}= \frac{{W}_{y_{(\alpha+1,\beta)}}-{W}_{y_{(\alpha-1,\beta)}}}{2\Delta x}-\frac{{W}_{x_{(\alpha,\beta+1)}}-{W}_{x_{(\alpha,\beta-1)}}}{2\Delta y}.
\end{equation}

\subsection*{Numerical calculation of the potential}

The potential is calculated by numerically solving the equations   $-\nabla V=\vy{W}$, taking into account that  $\nabla \times \vy{W} = 0$. For the
computation of the empirical potential, we used conditions $V$ = 0 in all the boundary regions of the box and then use the forward centered discretization formula for the gradient operator \cite{mazzoli2019field, hyman1997natural} starting from the city bounding box corner. In
a cell $i$ with indices ($\alpha$, $\beta$)
\begin{equation}
\frac{dV_i}{dx}=\frac{V_{(\alpha+1,\beta)}-V_{(\alpha,\beta)}}{\Delta x}={W}_{x_{(\alpha,\beta)}},
\end{equation}
and also
\begin{equation}
\frac{dV_i}{dy}=\frac{V_{(\alpha,\beta+1)}-V_{(\alpha,\beta)}}{\Delta y}={W}_{y_{(\alpha,\beta)}},
\end{equation}
the procedure is iterated until all cells have been assigned a potential. To decrease the noise, we average the resulting potentials starting from the four corners of the bounding box .

\subsection*{Parameter estimation}

In our d-mix model, the single free parameter is $\ell_c$, which directly determines whether the agents optimize or not the order of the trajectory stops.
For a given empirical dataset, we rely on $\rho$ to estimate  $\ell_c$. To accomplish this,  we define the following function
\begin{equation}
E(\ell_c)= \left| \rho_{data}- \rho( \ell_c) \right|,  
\end{equation}
where $\rho_{data}$ is obtained from the dataset for all trajectories, and $\rho(\ell_c)$ is calculated through the d-mix model with parameter $\ell_c$. The objective function can be minimized to yield an estimated value of $\ell_c$ needed to reproduce with d-mix the signs of the set of empirical trajectories.

\section*{Data Availability}

The D1 dataset on truck trajectories in China can be downloaded from the National Road Freight Supervision and Service Platform (\url{https://www.gghypt.net/}). 

The D2 dataset on Foursquare check-ins in New York City was obtained from the details given in Ref.  \cite{bao2012location}.

The 2010 census divisions in the case of New York were downloaded from \url{https://www.census.gov/geo/mapsdata/maps/block/2010/}.

\section*{Code Availability}

The code used for this work is available at \url{https://github.com/orgs/erjianliu}.

\section*{Acknowledgements}

We would like to thank Xin Lu for a critical reading and useful suggestions on the manuscript. 
E.L. and J.J.R. acknowledge funding from MCIN/AEI/10.13039/501100011033/FEDER/EU under project APASOS (PID2021-122256NB-C22) and from MCIN/AEI/10.13039/501100011033 under project Next4Mob (PLEC2021-007824) and the Maria de Maeztu Program for units of Excellence in R$\&$D CEX2021-001164-M.  X.-Y.Y. was supported by the National Natural Science Foundation of China (Grant No. 72271019).


%

\appendix

\renewcommand{\thefigure}{S\arabic{figure}}
\setcounter{figure}{0}
\renewcommand{\thetable}{S\arabic{table}}
\setcounter{table}{0}

\section{Mobility data}

\subsection*{Logistic data}
The logistic data (D1) we used comes from truck travel records of 21 Chinese cities, which can be
downloaded from the National Road Freight Supervision and Service Platform (https://www.gghypt.net/). 
The D1 dataset contains more than 2.7 million truck travel records, spanning from May 18, 2018 to May 31, 2018. Each truck travel records consists of  truck ID, longitude, latitude, speed, timestamp, see Table \ref{table:tables1}. In previous works data treatment was performed in order to identify trip ends and obtain track sequence according to the identified trip ends \cite{yang2022identifying, yang2022identifyingE}. For a truck’s base point, it could be a logistics center, or a freight hub. A truck frequently returns to the base point, due to the driver ending their work shift or reloading to start the next round of freight distribution. In view of this, we consider the most visited location by a truck as its
base point (origin) and extract the truck delivery route from the track sequence.

\begin{table*}
\begin{center}
\begin{tabular}{l*{6}{c}r}
ID  & Longitude & Latitude  & Speed (km/h) & Timestamp \\ 
\hline
60817be2749c77 & 119.786484 & 34.387562 &	0 &  2018-05-19 12:03:20  \\ 
60817be2749c77 & 119.787315 & 34.388016 &	30 & 2018-05-19 12:03:50 \\
60817be2749c77  & 119.788536 &	34.388783 &	25 & 2018-05-19 12:04:20 \\
60817be2749c77 & 119.789902	& 34.38847 &	7 & 2018-05-19 12:04:50\\
60817be2749c77	& 119.789902 &	34.38847 &	0 & 2018-05-19 12:05:20 \\

\end{tabular}
\caption{List of examples of heavy truck GPS data.}
\label{table:tables1}
\end{center}
\end{table*}

\begin{figure}
\centering
\includegraphics[width=8cm]{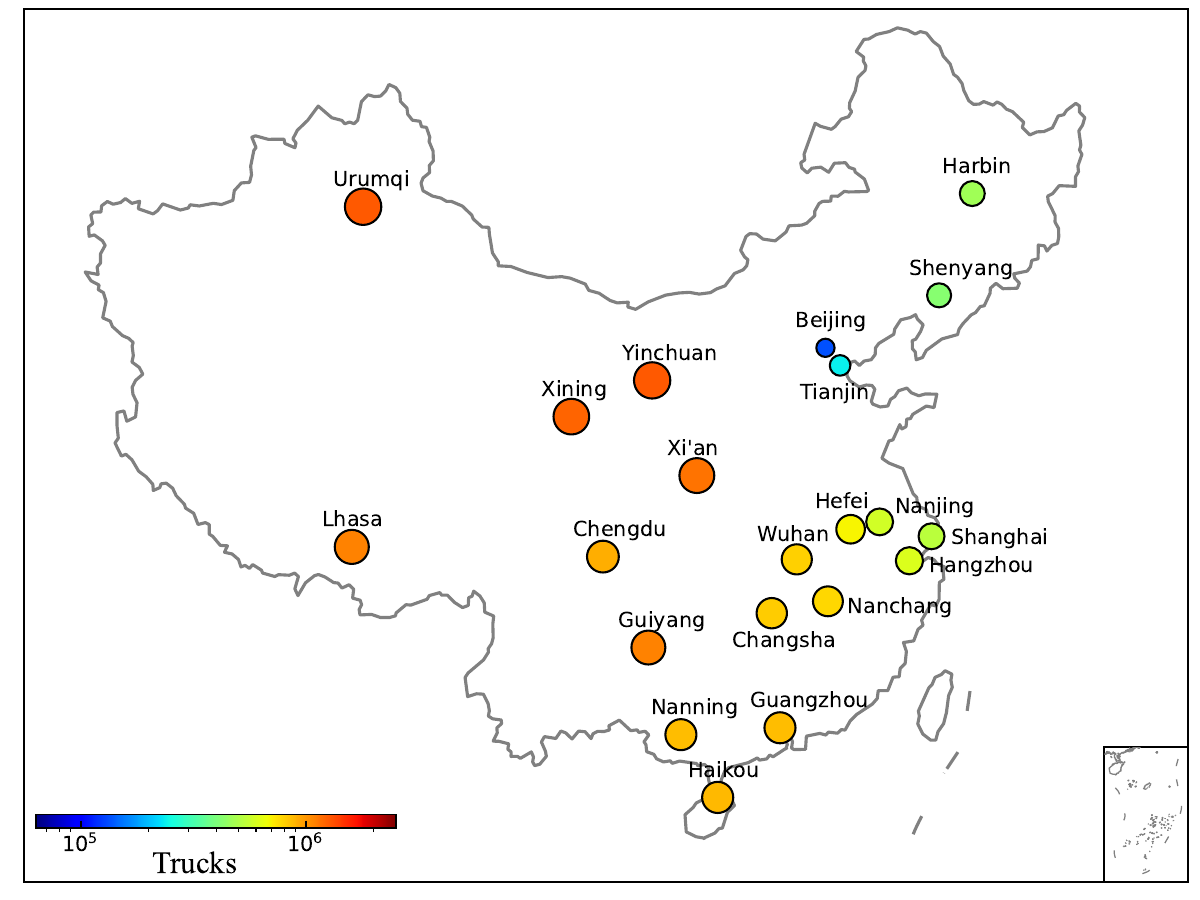}
\caption[Considered cities in logistic data.]{{\textbf {Considered cities in logistic data.}} The locations of the 21 cities and the number of trucks in each city.}
\label{figs1}
\end{figure}

As shown in Fig. \ref{figs1}, the 21 cities in D1 the most densely populated cities (e.g., Beijing and Shanghai) and less densely populated cities (e.g., Xining). These cities have different sizes, economies, cultural backgrounds and infrastructural setting. This gives us an important opportunity to study mobility patterns in heterogeneous social contexts. In Table \ref{table:tables2}, we list the trajectories, city center, population and GDP of the 21 cities we consider.

\begin{table*}
\begin{center}
\begin{tabular}{l*{10}{c}r}

City  & Trajectories & Latitude & Longitude & GDP ($\times 10^6$) & Population ($\times10^4$)   \\ 
\hline
Beijing &123036 &39.913818 & 116.363625 & 30320.00 & 2154.2   \\ 
Tianjin &104973 & 39.133331 & 117.183334 & 18809.64 & 1559.6  \\
Shanghai  & 180293 & 31.230400 & 121.473000 &  32679.87 &	2423.78  \\
Nanjing &51554 & 32.049999 & 118.766670 &  12820.40	& 843.62\\
Hangzhou	&58408 & 	30.250000 & 120.166664 &  16509.20 &	980.60  \\
Hefei &  582171 & 31.820591 & 117.227219 & 7822.91 & 808.70   \\ 
Nanchang & 54969 & 28.682892 & 115.858197 &  5274.67 & 554.55  \\
Changsha  &87979 & 	28.228209 & 112.938814 &  11003.41 &	815.47  \\
Wuhan & 85371 & 30.592849 & 114.305539 & 14847.29	& 1108.10\\
Yinchuan &20514 & 38.487193 & 106.230908 &  1901.48 &	225.06  \\
Urumqi  &14282 & 43.825592 & 87.616848 &  3099.77	& 350.58\\
Guangzhou & 46759 & 	23.129110 & 113.264385 &  22859.35 &	1490.44  \\
Lhasa &1906 & 	29.654838 & 91.140552 &  540.78 & 55.44   \\ 
Haikou &5942& 20.044412 & 110.198286 &  1510.51 & 230.23  \\
Nanning  &47949 &22.817002 & 108.366543 &  4009.00 &	666.16  \\
Chengdu &45121 & 30.657000 & 104.066002 &  15342.77	& 1633.00\\
Guiyang	&13164 & 	26.647661 & 106.630153 &  3798.45 &	488.19  \\
Shenyang & 62459 & 	41.805699 & 123.431472 &  6292.40 & 831.60  \\
Xi'an  &65126 & 	34.341574 & 108.939770 &  8349.86 &	1000.37  \\
Harbin &52235 & 45.803775 & 126.534967 &  6300.50	& 951.50\\
Xining	& 7892& 36.617134 & 101.778223 &  1286.41 &	237.11  \\

\end{tabular}
\caption{Number of trajectories, city center, population and GDP of the 21 cities.}
\label{table:tables2}
\end{center}
\end{table*}

\subsection*{Foursquare check-ins data}

\begin{figure}
\centering
\includegraphics[width=8cm]{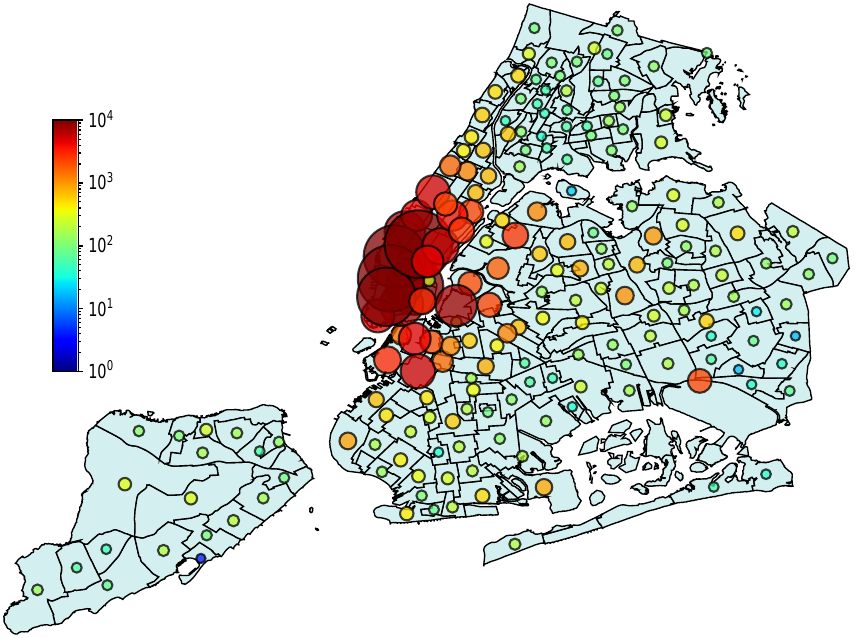}
\caption[Foursquare users check-ins in New York City.]{{\textbf {Foursquare users check-ins in New York City.} } Check-in times and zones distribution in NYC.}
\label{figs2}
\end{figure}
The Foursquare check-ins data (D2) is from  New York City (NYC).
Foursquare is a location-based social network on which users share their coordinates when checking in.
The D2 dataset contains  42035 individuals, in which 23520 users register trips among different analysis zones (here the space is divided according 2010 census areas, see https://www.census.gov/geo/maps
data/maps/block/2010/). We consider the most common check-in zone as the individual's trajectory origin. 
Similarly to what described when dealing with logistic data, we extract a sequence of consecutively visited locations with the last location corresponding to the initial one \cite{bao2012location}. Check-in times and zones distribution in NYC, see  Fig. \ref{figs2}

\section{The statistical description of empirical trajectories}

The city centers of the 21 Chinese cities for the stops distribution and flux calculations have been taken at
locations shown in  Table \ref{table:tables2}. For NYC we take the city center as (40.730610 \ N, -73.935242\ E).

\begin{figure}
\centering
\includegraphics[width=8cm]{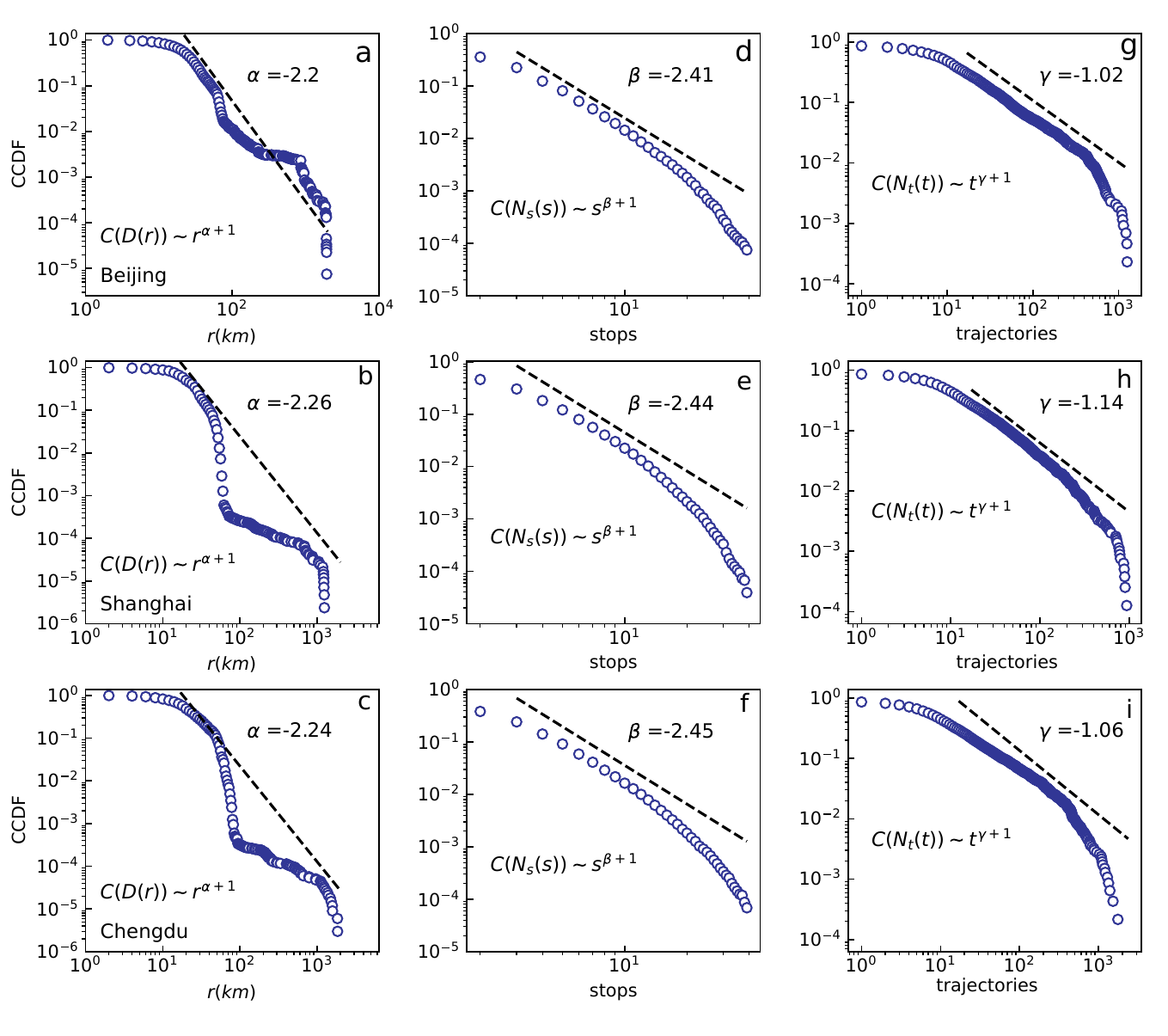}
\caption[Statistical description of Beijing, Shanghai and Chengdu trajectories]{{\textbf {The statistical description of Beijing, Shanghai, Chengdu.}} The complementary cumulative probability distributions (CCDF) of :
{\bf (a,d,g)} the distance between the trajectory stops and the city center;  {\bf (b,e,h)} number of stops per trajectory;  {\bf (c,f,i)} number of truck trajectories with the same origin.
The lines in each panel represent a power-law fit to the real data. The dashed lines are only illustrative, to give an impression of a simple function approaching the data to be used in the models, and not a product of fits. The trajectories considered for the statistics are those whose origin lays in a circle centered in three cities and with a radius of 20 km.}
\label{figs3}
\end{figure}

\begin{figure}
\centering
\includegraphics[width=8cm]{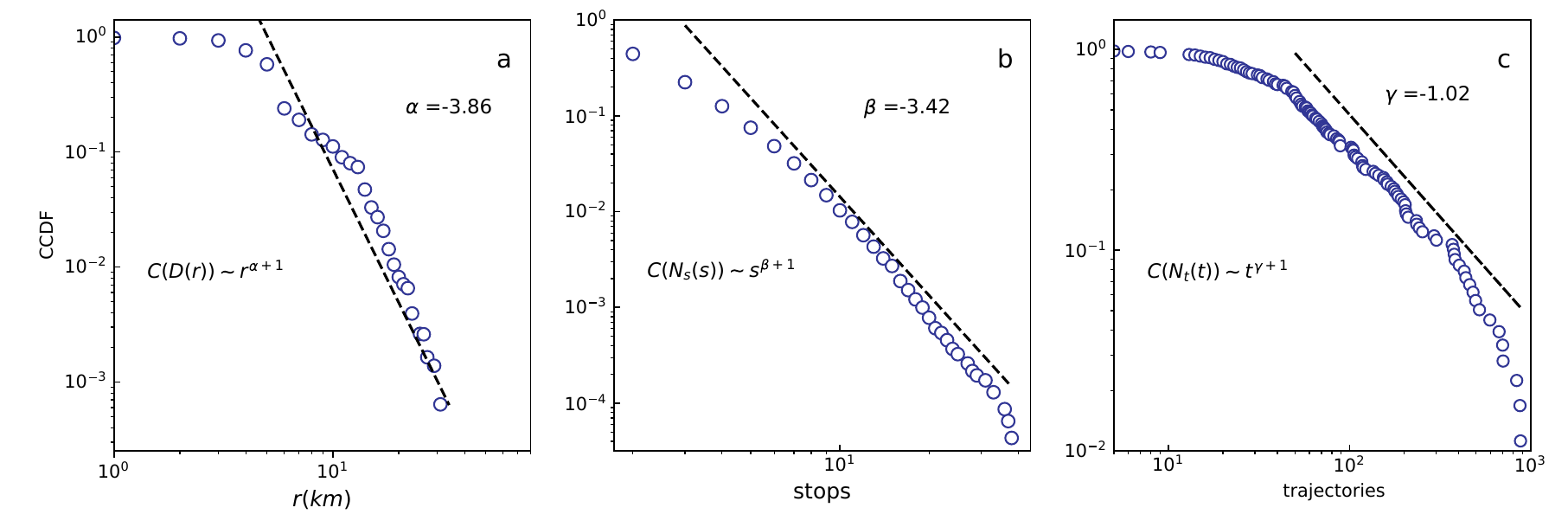}
\caption[Statistical description of New York trajectories]{{\textbf {The statistical description of New York.}} The complementary cumulative probability distributions (CCDF) of:
{\bf a}   the distance between the trajectory stops and the city
center;  {\bf b}   number of stops per trajectory;  {\bf c} number of truck trajectories with the same origin.
The lines in each panel represent a power-law fit to the real data. The dashed lines are only illustrative, to give an impression of a simple function approaching the data to be used in the models, and not a product of fits. The trajectories considered for the statistics are those whose origin lays in a circle centered in three cities and with a radius of 5 km.}
\label{figs4}
\end{figure}

The statistical description of the  trajectories in Beijing, Shanghai and Chengdu surrounding 20 km radius area is provided in Fig. \ref{figs3}.
The same description for NYC (surrounding 5 km radius area) is provided in Fig. \ref{figs4}.  In Table \ref{table:tables3}, we list the power-law exponents of Beijing, Shanghai, Chengdu and NYC.

\begin{table}
\begin{center}
\begin{tabular}{l*{6}{c}r}
City  & $\alpha$ & $ \beta$   & $\gamma$ \\ 
\hline
Beijing & 2.20 & 2.41 &	1.02  \\ 
Shanghai & 2.26 & 2.44 &1.14 \\
Chengdu  & 2.24 & 2.45 & 1.06 \\
NYC & 3.86 & 2.42 &	1.02  \\ 

\end{tabular}
\caption{List of power-law exponents of Beijing, Shanghai, Chengdu and NYC for the three distributions considered.}
\label{table:tables3}
\end{center}
\end{table}

\section{The geographical city center as RP}
In the main text, we use the origin of each trajectory as the RP.
Here we show the results when we consider the
geographical center of the city as RP.
As shown in Figs. S5-S7,  the fraction of positive $H(t) > 0$, negative $H(t) < 0$ and balanced $H(t) = 0$ trajectories with $3$, $4$, $5$, $6$ stops and for all trajectories of trucks serving the Chinese cities. 

Compared to use the origin of each trajectory as the RP, we can find that trajectories orientation across cities is significantly different and the values of $\rho$ for all trajectories is  related to the urban population.
This is probably because when we use the city center as RP, some details of the city are taken into account, such as the city shape, distribution of stops, streets and communication axes (e.g. highways).

\begin{figure}
\centering
\includegraphics[width=8cm]{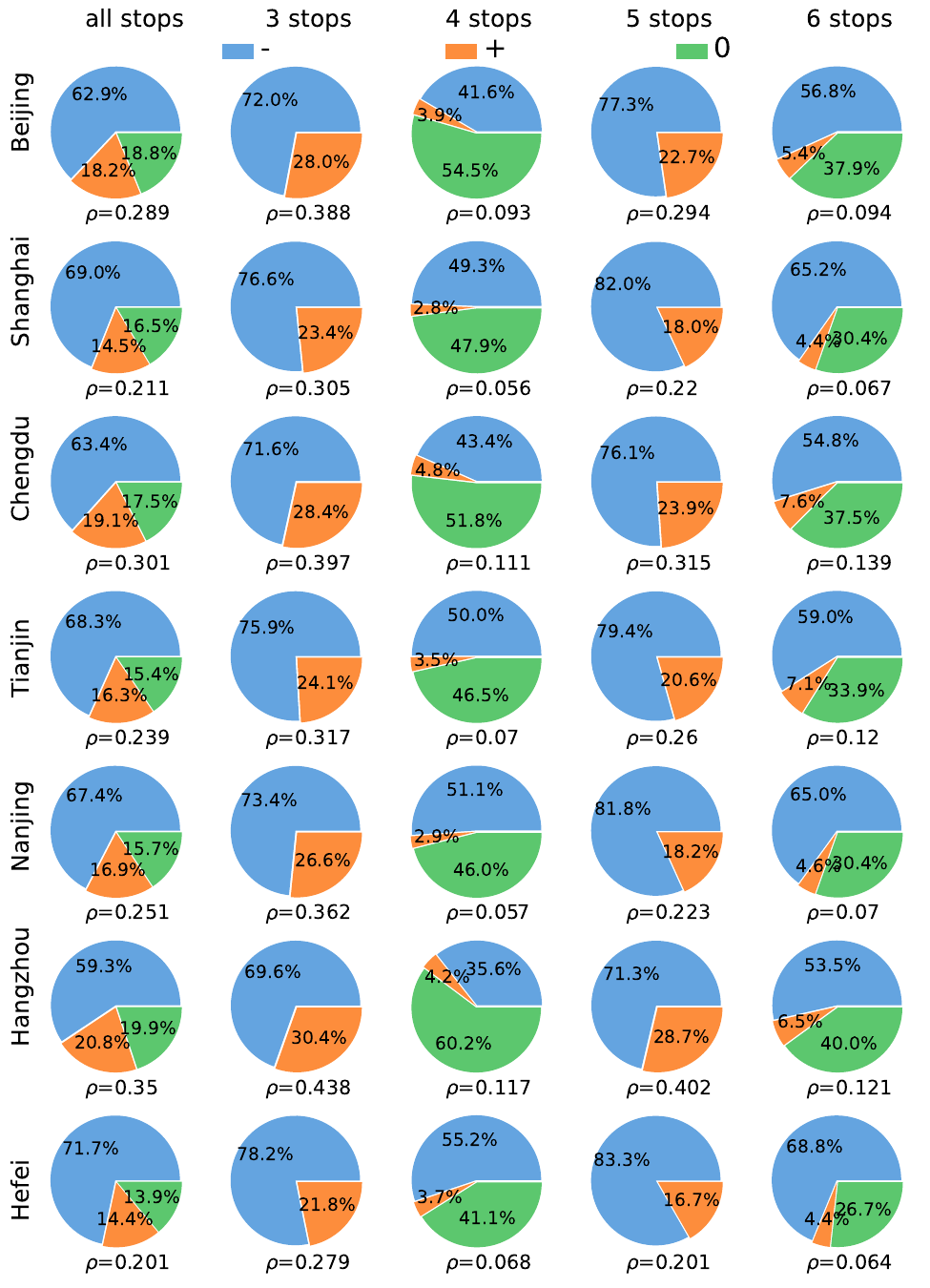}
\caption[Unbalance of empirical trajectories  using the city center as RP]{{\textbf {Unbalance of empirical trajectories of  using
geographical center of the city as RP.} }  
Fraction of positive $H(t) > 0$, negative $H(t) < 0$ and balanced $H(t) = 0$ trajectories with $3$, $4$, $5$, $6$ stops and for all trajectories of trucks serving Beijing, Shanghai, Chengdu, Tianjin, Nanjing, Hangzhou  and Hefei.}
\label{figs5}
\end{figure}

\begin{figure}
\centering
\includegraphics[width=8cm]{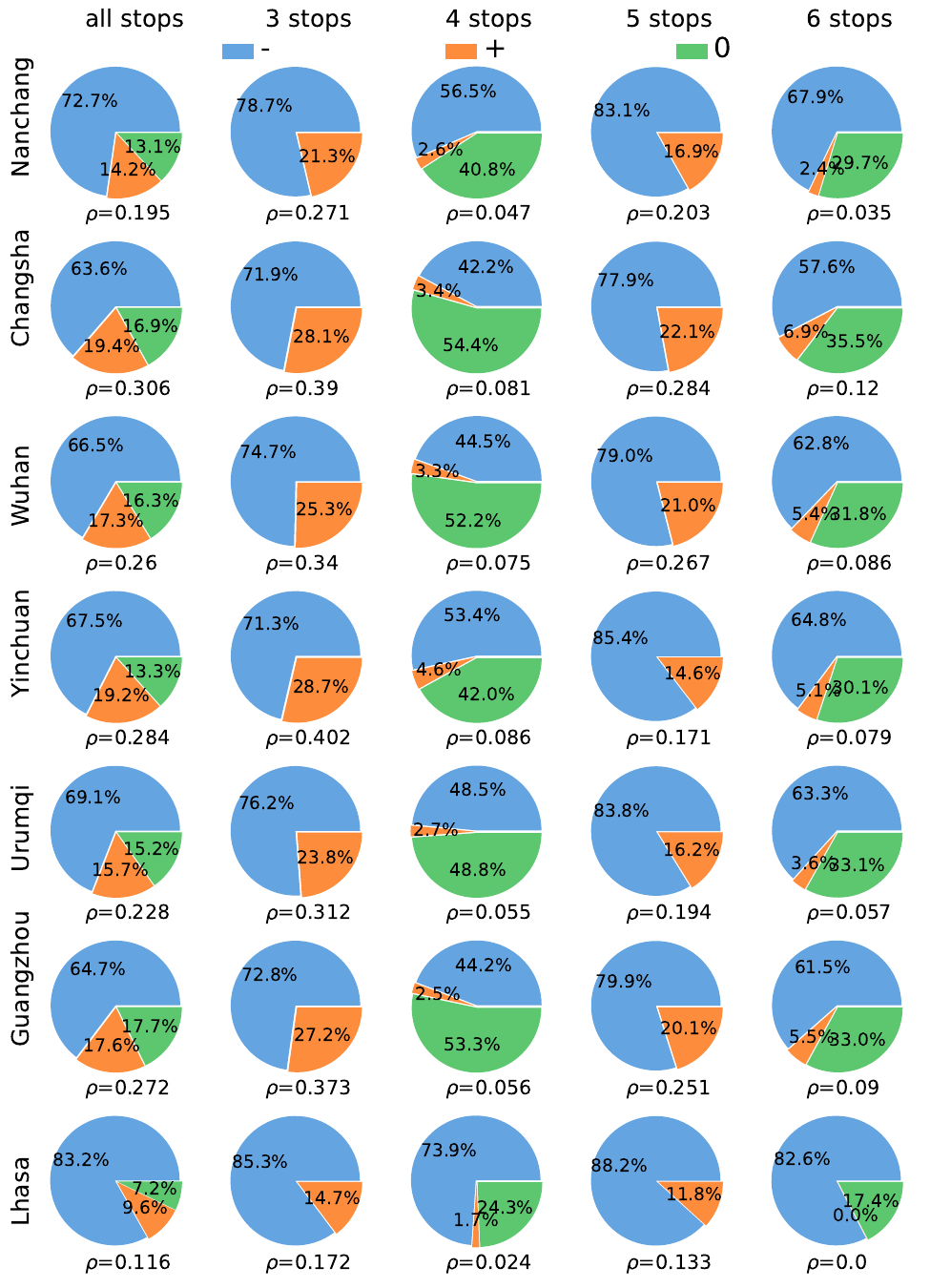}
\caption[Unbalance of empirical trajectories  using the city center as RP]{{\textbf {Unbalance of empirical trajectories of  using
geographical center of the city as RP.} }  
Fraction of positive $H(t) > 0$, negative $H(t) < 0$ and balanced $H(t) = 0$ trajectories with $3$, $4$, $5$, $6$ stops and for all trajectories of trucks serving Nanchang, Changsha, Wuhan, Yinchuan, Urumqi, Guangzhou  and Lhasa.}
\label{figs6}
\end{figure}

\begin{figure}
\centering
\includegraphics[width=8cm]{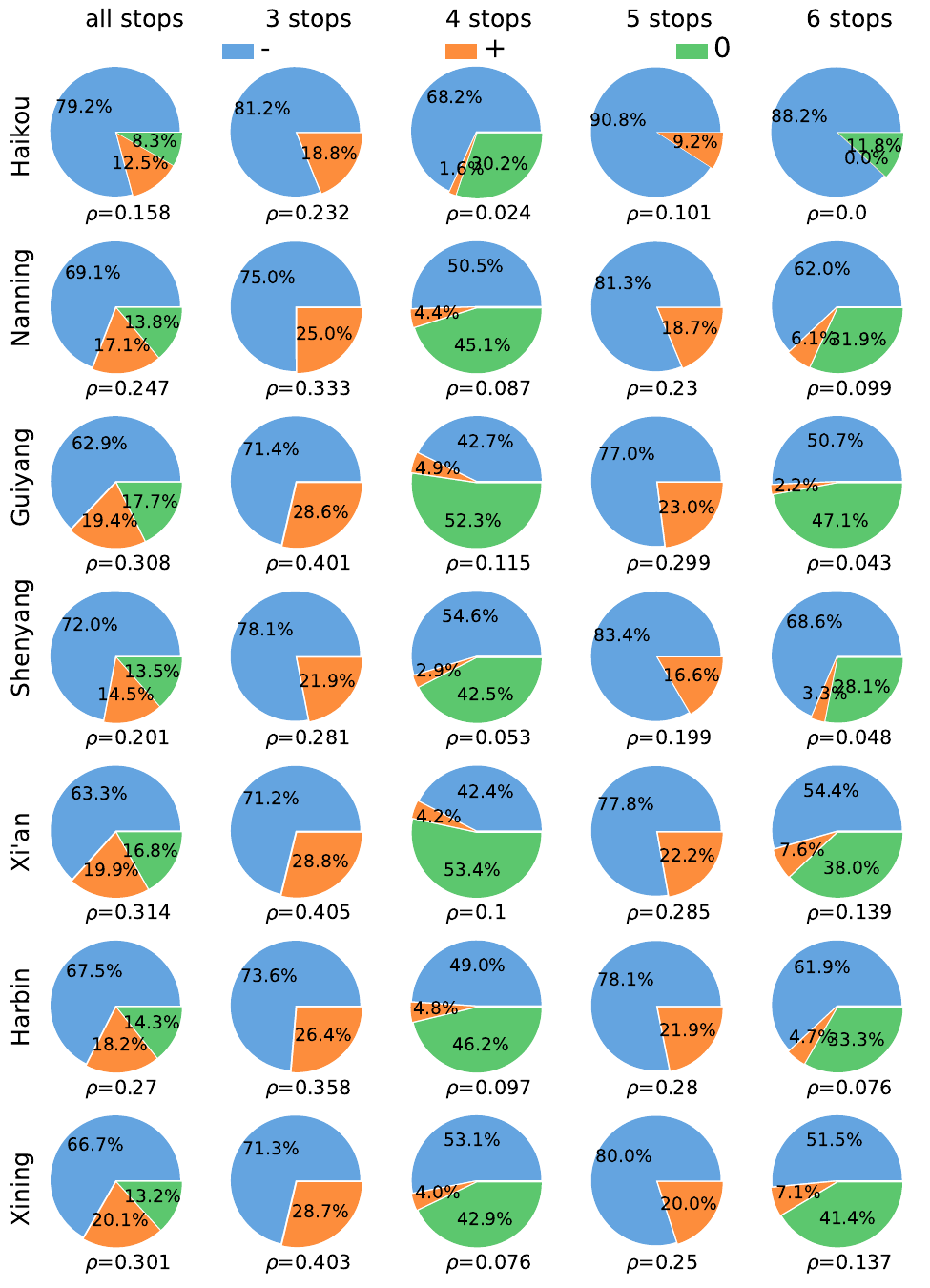}
\caption[Unbalance of empirical trajectories  using the city center as RP]{{\textbf {Unbalance of empirical trajectories of  using
geographical center of the city as RP.} }  
Fraction of positive $H(t) > 0$, negative $H(t) < 0$ and balanced $H(t) = 0$ trajectories with $3$, $4$, $5$, $6$ stops and for all trajectories of trucks serving Haikou, Nanning, Guiyang, Shenyang, Xi'an, Harbin and Xining. }
\label{figs57}
\end{figure}


\section{Relationship between $\rho$ and the field }

The unbalance factor $\rho$ is defined based on the trajectories. Recalling the form of $\rho$:
\begin{equation}
\rho = \frac{N_+}{N_-},
\end{equation}
where $N_+$ and $N_-$ are the number of positive and negative trajectories, respectively. Besides signed trajectories, there are also balanced ones up to a  number $N_0$. The total number of trajectories is thus $N = N_++N_-+N_0$. 

The objective here is to prove that $\rho < 1$ implies the existence of a mobility field. Note that this is a necessary condition, but not a sufficient one, since as we will see a field can still exist even if $\rho = 1$. 
 
Every trajectory is composed by a sequence of $s$ stops $\{\vv{x}_0, \vv{x}_1, ..., \vv{x}_{s-1}\}$. Between each couple of consecutive stops $i$ and $j$, we have defined a unit vector $\vv{ij}$ that has a sign depending on whether they  point toward the RP or away. Each trajectory has $s$ vectors and its final sign is the majority sign of them. These unit vectors, summed up vectorially in each cell of the space grid, generate the field.  In order for a field to exist it is thus necessary to have an unbalance either global or local in the number of positive and negative vectors. In fact, we can define an auxiliary ratio 
\begin{equation}
\rho' = \frac{S_+}{S_-},
\end{equation}
where $S_+$ and $S_-$ are the global number of positive and negative, respectively, unit vectors. Note that $\rho' < 1 $ implies the  unbalance of the unit vectors and, therefore, a global mobility field. 

Given that each trajectory has generically  $s$
stops and, consequently, $s$ unit vectors, we will call $s_+$ and $s_-$, respectively, to the number of positive and negative vectors.  We can then write that $s = s_++s_-$ in each trajectory. When considering all the trajectories, we can define  $\langle s_+\rangle_+$ and $\langle s_-\rangle_+$ as the average number of positive and negative vectors in positive trajectories. The corresponding averages for negative trajectories are $\langle s_+\rangle_-$ and $\langle s_-\rangle_-$, and for balanced trajectories $\langle s_+\rangle_0$ and $\langle s_-\rangle_0$. With these averages, we can write $\rho'$ as 
\begin{equation}
\rho' = \frac{S_+}{S_-}= \frac{N_+ \, \langle s_+ \rangle_{+}+ N_- \, \langle s_+ \rangle_{-}+ N_0 \,  \langle s_+ \rangle_{0}}{N_+ \, \langle s_- \rangle_{+}+ N_- \,  \langle s_- \rangle_{-}+ N_0 \, \langle s_- \rangle_{0}} .
\end{equation}
Dividing above and below in the ratio by $N_-$, we find
\begin{equation}
\rho' =  \frac{\rho \, \langle s_+ \rangle_{+}+  \langle s_+ \rangle_{-}+ (N_0/N_-) \,  \langle s_+ \rangle_{0}}{\rho \, \langle s_- \rangle_{+}+   \langle s_- \rangle_{-}+ (N_0/N_-) \, \langle s_- \rangle_{0}} .
\end{equation}
Some conditions are clear by definition: The first is that $\langle s_+ \rangle_{+} > \langle s_- \rangle_{+}$ and $\langle s_- \rangle_{-} > \langle s_+ \rangle_{-}$. The second one is that $ \langle s_- \rangle_{0} =  \langle s_+ \rangle_{0} = s_0\, /2$, where $s_0$ is the average length of balanced trajectories. Then $\rho'$ can be written as
\begin{equation}
\label{rhop}
\rho' =  \frac{\rho \, \langle s_+ \rangle_{+}+  \langle s_+ \rangle_{-}+ (N_0/N_-) \,  (s_0/2)}{\rho \, \langle s_- \rangle_{+}+   \langle s_- \rangle_{-}+ (N_0/N_-) \, (s_0/2)} .
\end{equation}
Modifying this equation to have $\rho$ as a function of $\rho'$, we can impose the condition that $\rho < 1$ to obtain: 
\begin{equation}
\rho =  \frac{ \rho' \, \left( \langle s_- \rangle_{-} + (N_0/N_-)\, (s_0/2)
\right) - \langle s_+ \rangle_{-} - (N_0/N_-)\, (s_0/2)}{\langle s_+ \rangle_{+}- \rho'\, \langle s_- \rangle_{+}} <1.
\end{equation}
This implies that for $\rho < 1$, $ \rho'$ must be
\begin{equation}
\label{eq1}
\rho' <  \frac{ \langle s_+ \rangle_{+} + \langle s_+ \rangle_{-} + (N_0/N_-)\, (s_0/2)}{ \langle s_- \rangle_{-} +  \langle s_- \rangle_{+}+ (N_0/N_-)\, (s_0/2) } = \rho'_{\rm sup}.
\end{equation}
A table with the statistics of the data trajectories and for those of the d-mix model are provided in Table \ref{table:tables4}. Inputing this information into Eq. \eqref{eq1}, we obtain the values of $\rho'_{\rm sup}$ shown in Table \ref{table:tables5}. All of them are smaller or equal than one and, therefore, $\rho < 1$ implies $\rho' < 1$.

\begin{table*}
\begin{center}
\begin{tabular}{l*{8}{c}r}
City/model  & $\langle s_+ \rangle_{+}$ & $\langle s_- \rangle_{+}$    & $\langle s_+ \rangle_{-}$ & $\langle s_- \rangle_{-}$ & $\langle s\rangle_{0}$ & $\langle s\rangle_{+}$ & $\langle s \rangle_{-}$ & $N_0/N_- $  \\ 
\hline
Beijing & 2.32 & 1.20 &	1.23 & 2.79 & 2.14 &	3.52 & 4.02 & 0.25 	\\ 
Shanghai & 2.38 & 1.24 &	1.27 & 2.89 & 2.17&	3.62 & 4.16 & 0.24 \\ 
Chengdu  & 2.37 & 1.24 &	1.26 & 2.87 & 2.17&	3.61 & 4.13 & 0.25 	\\ 
New York & 2.47 & 1.32 &	1.39 & 3.06 & 2.3&	3.79 & 4.45 & 0.23 \\ 
d-rand model & 2.29 & 1.20 &	2.23 & 6.81 & 2.24&	3.49 & 9.04 & 0.10 \\
d-mix model & 5.26 & 3.58 &	3.00 & 5.85 & 3.18&	8.84 & 8.85 & 0.23 \\ 
d-TSP model & 5.26 & 3.54 &	3.06 & 5.78 & 3.23&	8.80 & 8.84 & 0.26 \\
\end{tabular}
\caption{List of the values measured from all datasets, d-mix, d-TSP and d-rand model.  } 
\label{table:tables4}
\end{center}
\end{table*}

\begin{figure}
\centering
\includegraphics[width=8cm]{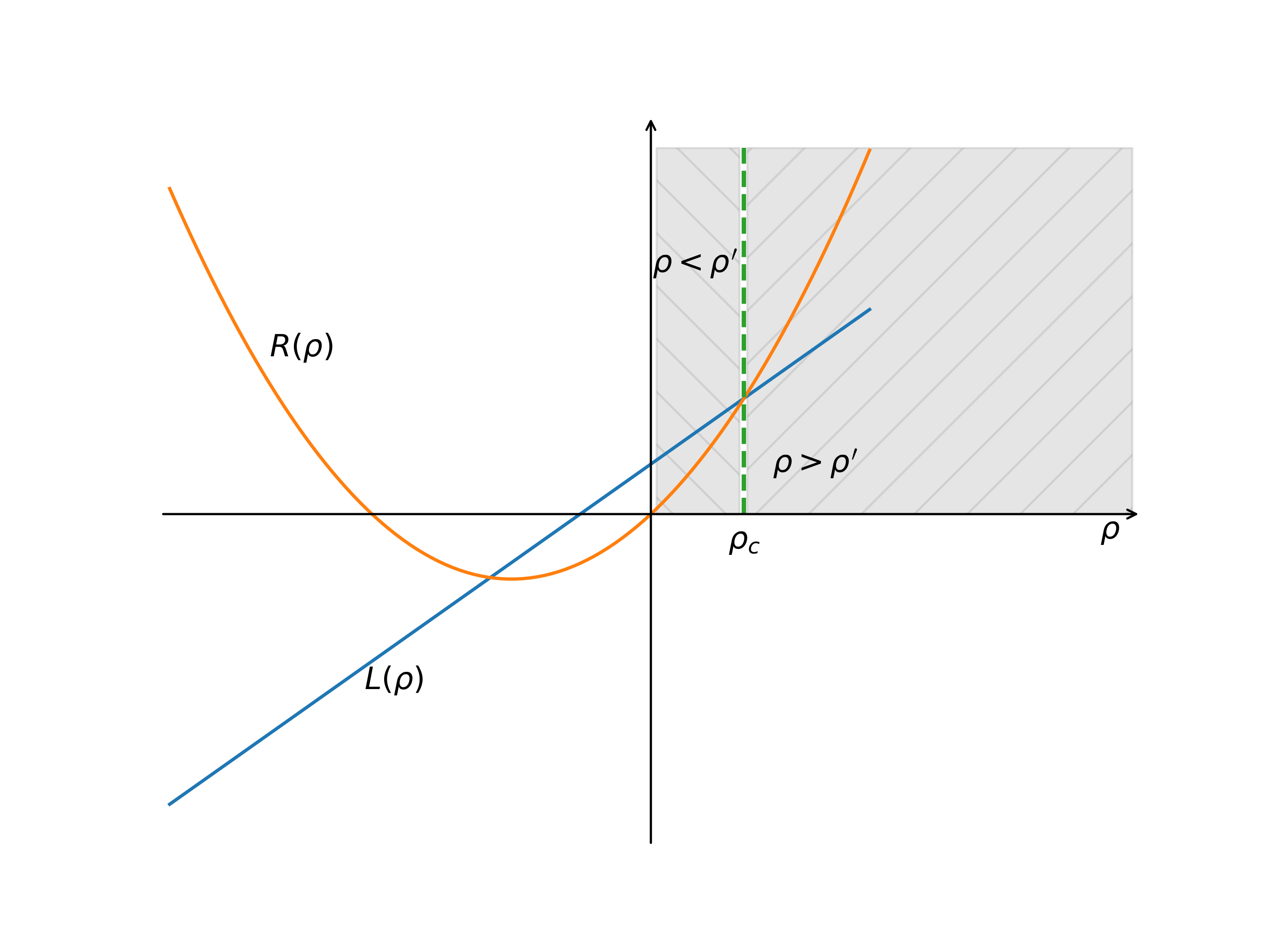}
\caption[Critical value of $\rho$]{{\textbf {Critical value of $\rho$.}} Crossing point of the curves $R(\rho)$ and $L(\rho)$. This point marks the value of $\rho$ for which the relation $\rho' > \rho$ passes to $\rho' < \rho$.  }
\label{figs9}
\end{figure}

An alternative way of demonstrating this condition and  gaining some insights in the process on the relation between $\rho$ and $\rho'$ is to wonder when $\rho$ is smaller or larger than $\rho'$.  We retake Eq. \eqref{rhop} and impose that $\rho' > \rho$, this yields: 
\begin{align}
& \rho \, \langle s_+ \rangle_{+}+  \langle s_+ \rangle_{-}+ (N_0/N_-) \,  (s_0/2) > \nonumber \\ & \rho \, \left[ \rho \, \langle s_- \rangle_{+}+   \langle s_- \rangle_{-}+ (N_0/N_-) \, (s_0/2) \right].
\end{align}
The left side of the expression is a linear function $L(\rho)$ in the domain of $\rho$ values, while the right side $R(\rho)$ is a parabola so at certain point for a  value of $\rho$, $\rho_c$,  both curves cross ($L(\rho_c) = R(\rho_c)$, see Fig. \ref{figs9}). For $\rho < \rho_c$, we have $\rho' > \rho$, while we have $\rho' <  \rho$ for $\rho > \rho_c$. The crossing point occurs when 
\begin{align}
&\rho \, \langle s_+ \rangle_{+}+  \langle s_+ \rangle_{-}+ (N_0/N_-) \,  (s_0/2) = \nonumber\\ & \rho \, \left[ \rho \, \langle s_- \rangle_{+}+   \langle s_- \rangle_{-}+ (N_0/N_-) \, (s_0/2) \right].
\end{align}
Therefore, the expression for $\rho_c$ is 
\begin{widetext}
\begin{align}
&\rho_c = \\ 
& \frac{\langle s_+ \rangle_{+} - \langle s_- \rangle_{-}  - (N_0\, s_0)/(2\, N_-) +\sqrt{[ \langle s_- \rangle_{-} -  \langle s_+ \rangle_{+} + (N_0\, s_0)/(2\, N_-)]^2 + 4\, \langle s_- \rangle_{+} \, [\langle s_+ \rangle_{-}  + (N_0\, s_0)/(2\, N_-) ]}}{2\, \langle s_- \rangle_{+}}.  \nonumber    
\end{align}
\end{widetext}
The question is thus whether $\rho_c$ is larger or smaller than one. If $\rho_c <1$, then when $\rho$ approaches $1$ from below but it is larger than $\rho_c$ then we have that $\rho' < \rho <1$ and, therefore $\rho < 1$ implies as well that $\rho' < 1$. Table \ref{table:tables5} shows the values of $\rho_c$ for all the datasets and the  models. In all the cases, $\rho_c < 1$. Therefore in our case, the condition  $\rho < 1 \implies \rho' < 1$ holds. 

Note that this is not a general demonstration, it is only valid for our cases (empirical datasets and models). There can be values of the statistics of the trajectories for which this relation between $\rho$ and the field does not hold. This second demonstration is interesting because it could help to determine in more general cases the conditions for the existence of the field given the value of $\rho$ and the critical value $\rho_c$.

\begin{table}
\begin{center}
\begin{tabular}{l*{8}{c}r}
City/model   & $\rho_c$  & $\rho'_{sup} $\\ 
\hline
Beijing  & 0.85& 0.88\\ 
Shanghai  & 0.84 & 0.88\\ 
 Chengdu & 0.84& 0.88\\ 
 New York& 0.84& 0.85\\ 
 d-rand model& 0.60& 0.40\\
 d-mix model& 0.83& 1.00\\ 
 d-TSP model& 0.87& 1.00\\

\end{tabular}
\caption{List of the values of $\rho_c$ and $\rho'_{sup}$   for all the datasets, d-mix, d-TSP and d-rand model.}
\label{table:tables5}
\end{center}
\end{table}



\section{Unbalance of empirical trajectories in NYC and further Chinese cities}
\begin{figure}
\centering
\includegraphics[width=8cm]{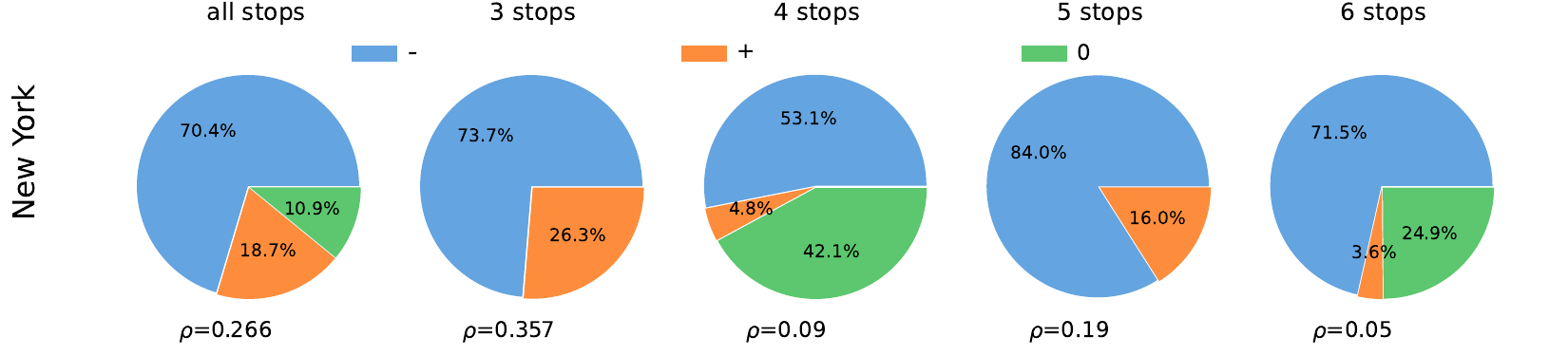}
\caption[Unbalance of empirical trajectories in NYC using the trajectory origin as RP]{{\textbf {Unbalance of empirical trajectories in New York City.}} 
Fraction of positive $H(t) > 0$, negative $H(t) < 0$ and balanced $H(t) = 0$ trajectories with $3$, $4$, $5$, $6$ stops and for all trajectories in New York, using the trajectory origin as RP. 
}
\label{fig10}
\end{figure}

In the main text, we have shown the unbalance of empirical trajectories in Beijing, Shanghai and Chengdu. To show our results are robust, here we show the fraction of positive $H(t) > 0$, negative $H(t) < 0$ and balanced $H(t) = 0$ trajectories with $3$, $4$, $5$, $6$ stops and for all trajectories of individuals in NYC  (see Fig. \ref{fig10}) and trajectories of trucks in more Chinese cities (the other 18 Chinese cities),  using the trajectory origin as RP, see Figs. S10-S12.

Combining the results of the main text, we can find that
the fraction of $N_+$ and $N_-$ for trajectories of a fixed number of stops and any number of stops is very close across cities. The number of trajectories with negative orientation $N_-$ is dominant, hence $\rho < 1$ in all cities. 

Note that Foursquare check-ins data encode a different type of mobility compared to logistic data. The values of $\rho$ of individuals' check-ins are different from the one obtained from  logistic data both for trajectories of a certain stops number only or for all trajectories.

\begin{figure}
\centering
\includegraphics[width=8cm]{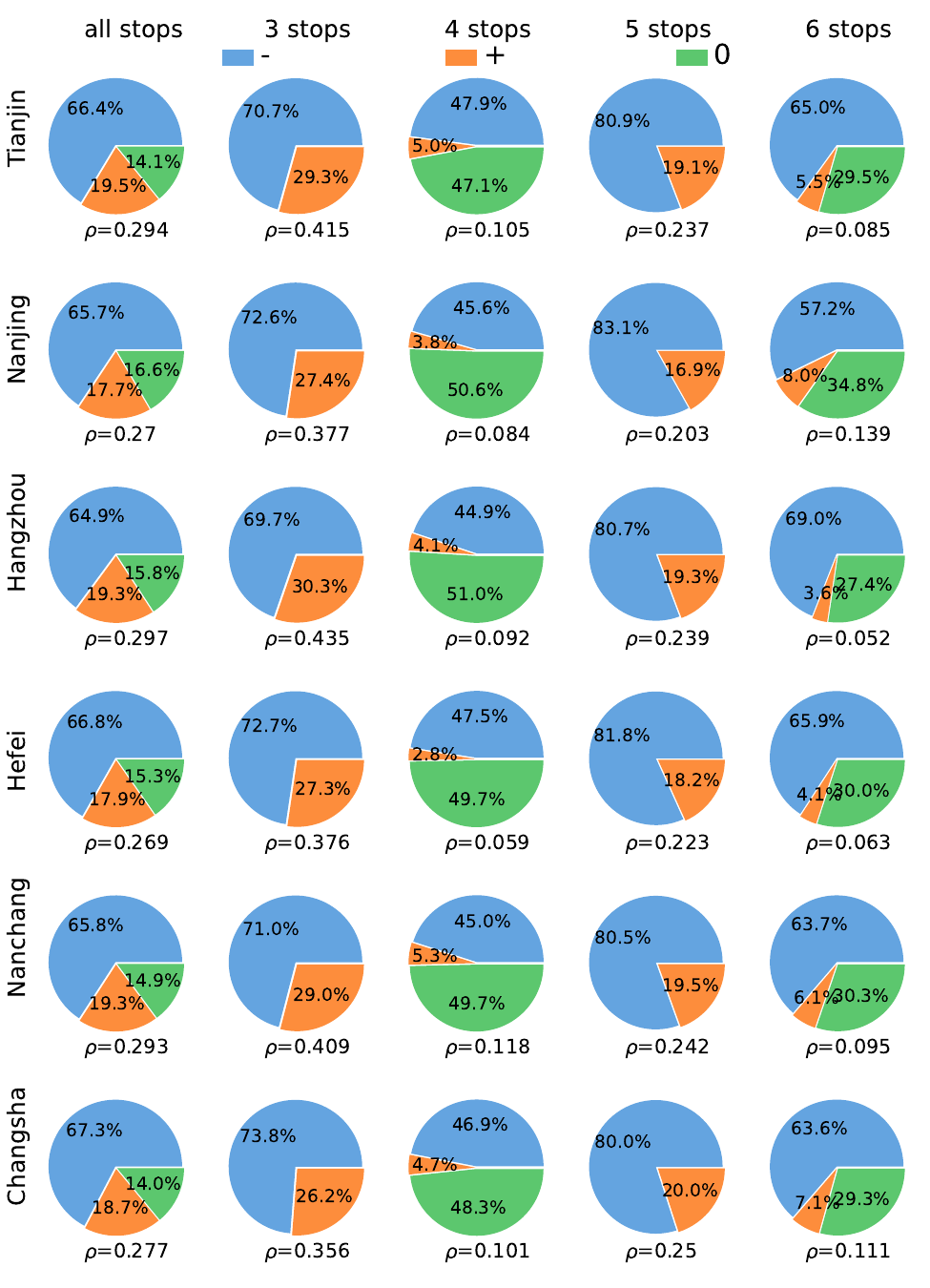}
\caption[Unbalance of empirical trajectories in Chinese cities]{{\textbf {Unbalance of empirical trajectories in Chinese cities.}} Fraction of positive $H(t) > 0$, negative $\$H(t) < 0$ and balanced $H(t) = 0$ trajectories with $3$, $4$, $5$, $6$ stops and for all trajectories of trucks serving Tianjin, Nanjing, Hangzhou, Hefei, Nanchang and Changsha, using the trajectory origin as RP. }
\label{figs11}
\end{figure}

\begin{figure}
\centering
\includegraphics[width=8cm]{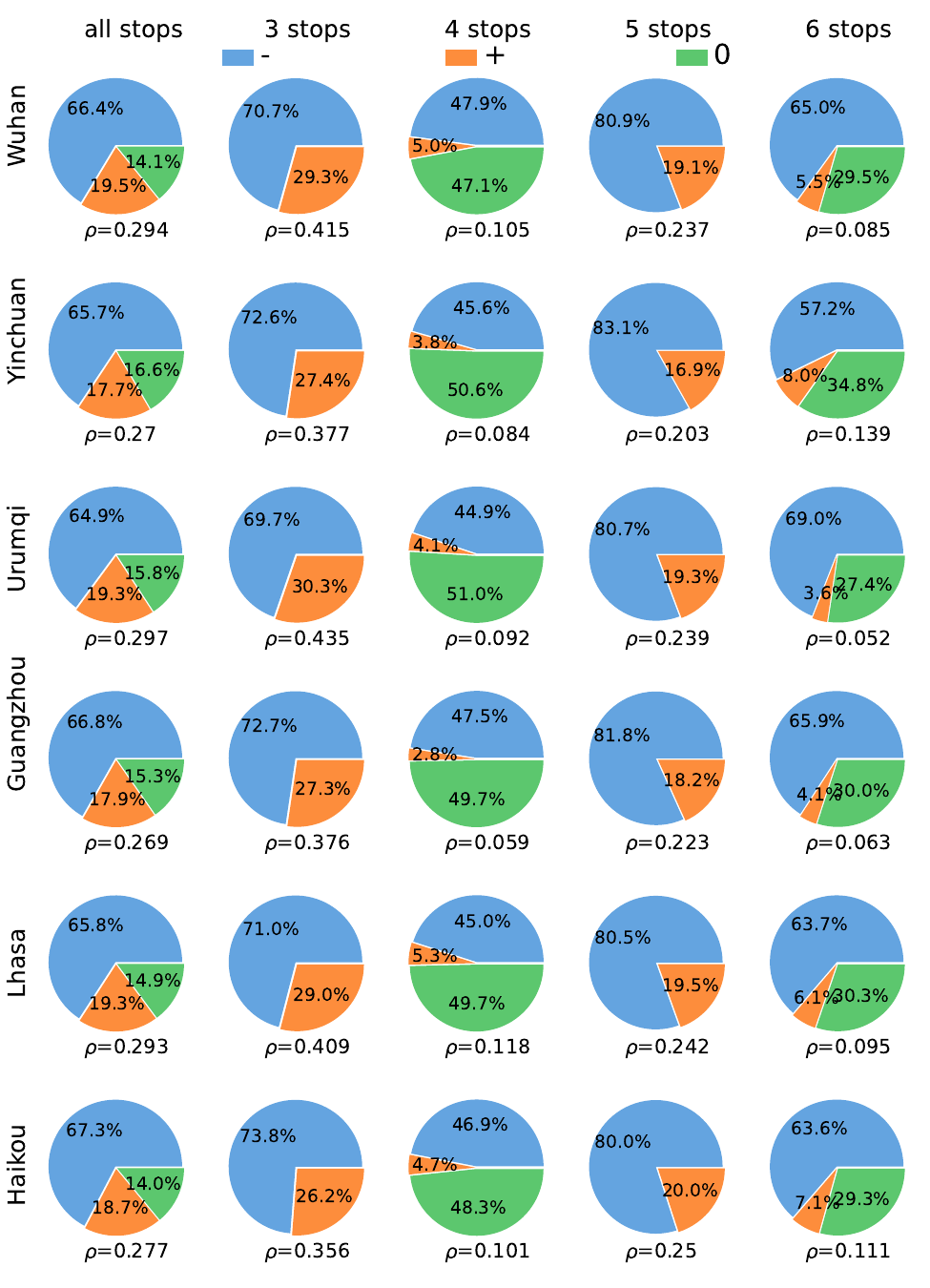}
\caption[Unbalance of empirical trajectories  using the trajectory origin as RP]{{\textbf {Unbalance of empirical trajectories in Chinese cities.}} Fraction of positive $H(t) > 0$, negative {$H(t) < 0$} and balanced {$H(t) = 0$} trajectories with $3$, $4$, $5$, $6$ stops and for all trajectories of trucks serving Wuhan, Yinchuan, Urumqi, Guangzhou, Lhasa and Haikou, using the trajectory origin as RP. }
\label{figs12}
\end{figure}

\begin{figure}
\centering
\includegraphics[width=8cm]{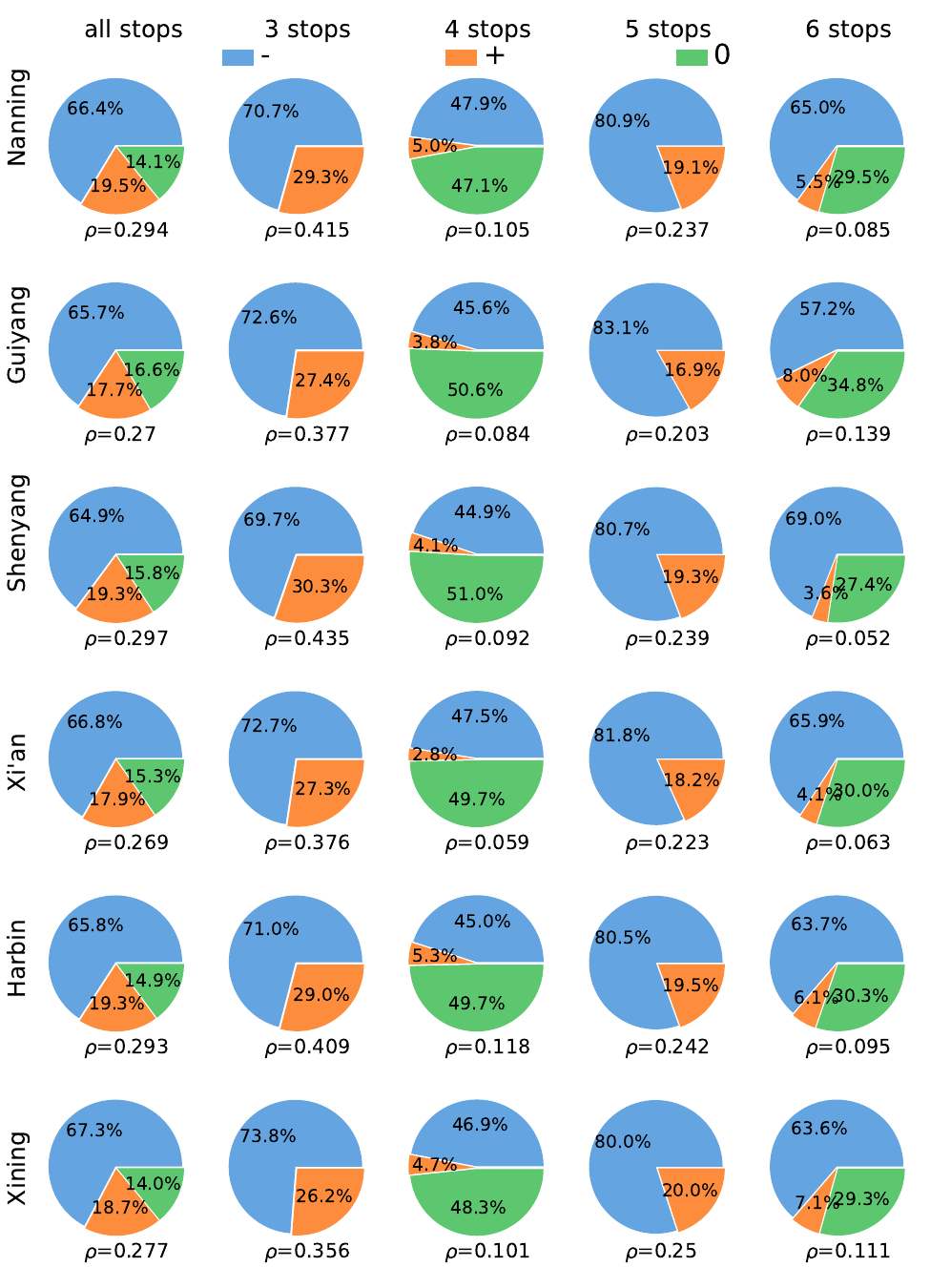}
\caption[Unbalance of empirical trajectories  using the trajectory origin as RP]{{\textbf {Unbalance of empirical trajectories in Chinese cities.}} Fraction of positive $H(t) > 0$, negative $H(t) < 0$ and balanced $H(t) = 0$ trajectories with $3$, $4$, $5$, $6$ stops and for all trajectories of trucks serving Nanning, Guiyang, Shenyang, Xi'an, Harbin and Xining, using the trajectory origin as RP. }
\label{figs13}
\end{figure}

\section{Rand model flux results}

In the main text, we have defined the Rand model.
Here we run the Rand model to generate trajectories in two different scenarios. The first scenario is a circular city of radius $R$ within a square of side $L$, the second one is a circular city of radius $R$ within a space with periodic boundary conditions. We check the features of Rand model flux as a function of the distance to the city center, see Figs. S13-S14.
With open boundary conditions, we observe that the flux decreases (it is more negative) as we get closer to the bounding box. However, with periodic boundary conditions, we find that the flux is fluctuating around zero and, therefore, the vectors are random in direction and module.


\begin{figure}
\centering
\includegraphics[width=8cm]{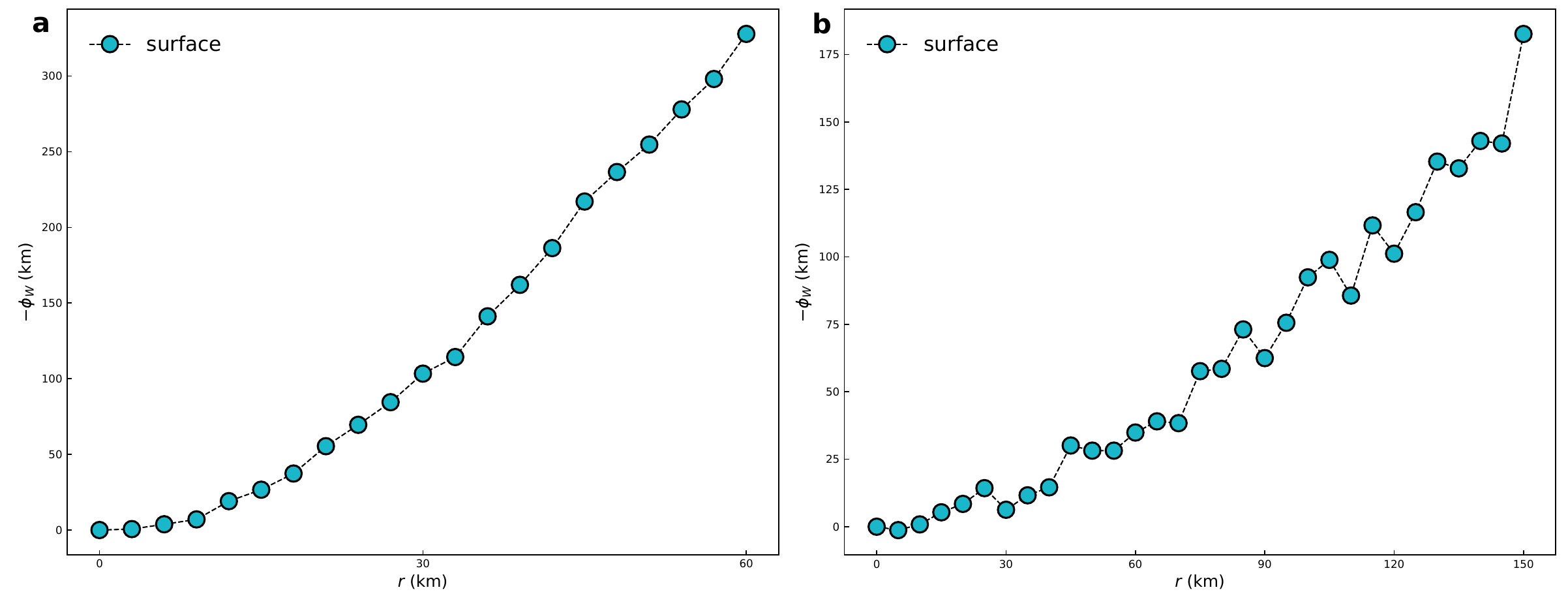}
\caption[Properties of flux with open boundary conditions]{{\textbf {Properties of the flux of Rand model with open  boundary conditions.}} 
 Here we set the values of $R = 20 \, km$ in a space limited by a box of side $L = 100 \, km$ in \textbf{a} and $L = 400 \, km$ in \textbf{b} with $100\, 000$ trajectories. The flux is displacing to the right as the value of the bounding box frame $L$ increases.}
\label{fig104}
\end{figure}

\begin{figure}
\centering
\includegraphics[width=8cm]{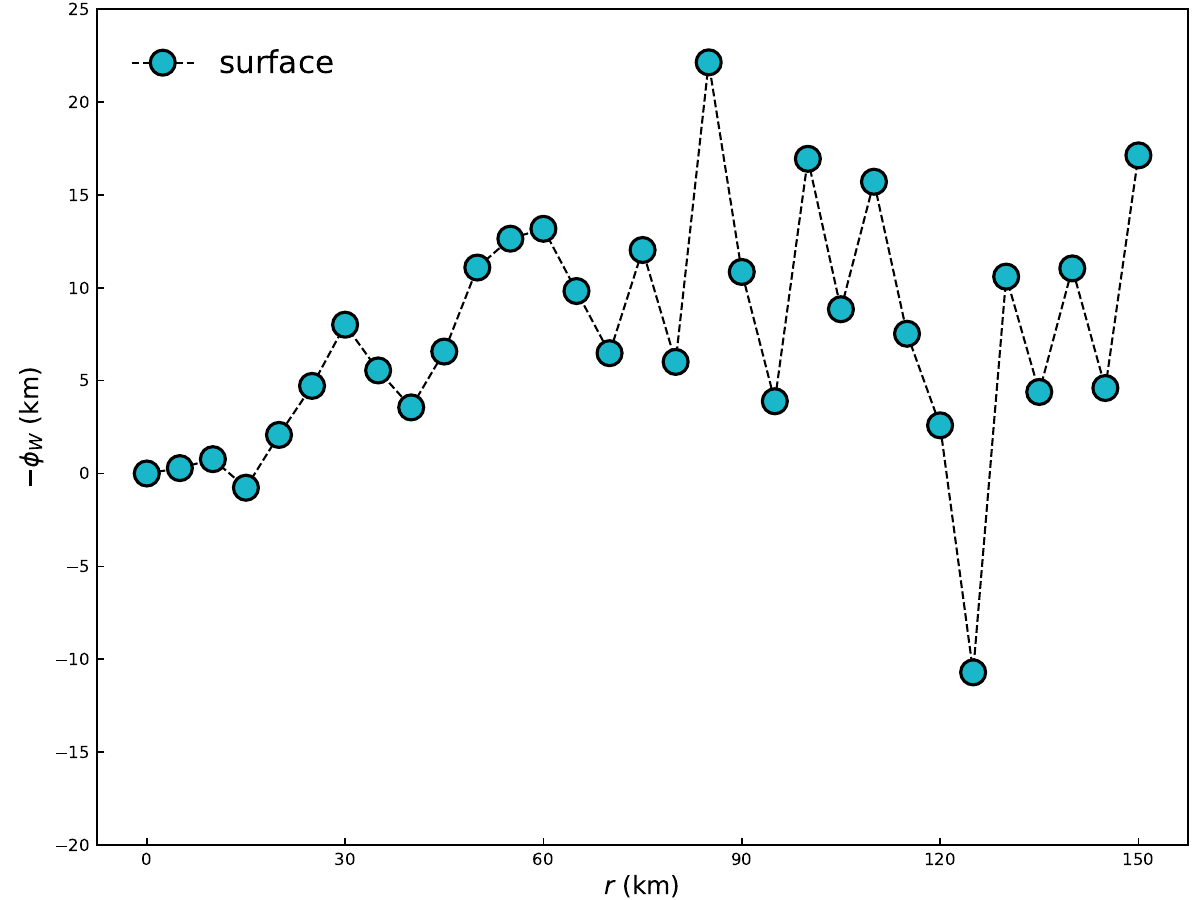}
\caption[Properties of the flux of Rand model with periodic boundary conditions]{{\textbf {Properties of the flux of Rand model with  periodic boundary conditions.}} 
Here we set
the values of $R = 20 \, km$ in a space  limited by a box of side $L = 400 \, km$ periodic boundary conditions with $100\, 000$ trajectories. }
\label{fig105}
\end{figure}

\section{Sensitivity analysis for the d-mix model for $R$ and $L$. }
\begin{figure}
\centering
\includegraphics[width=8cm]{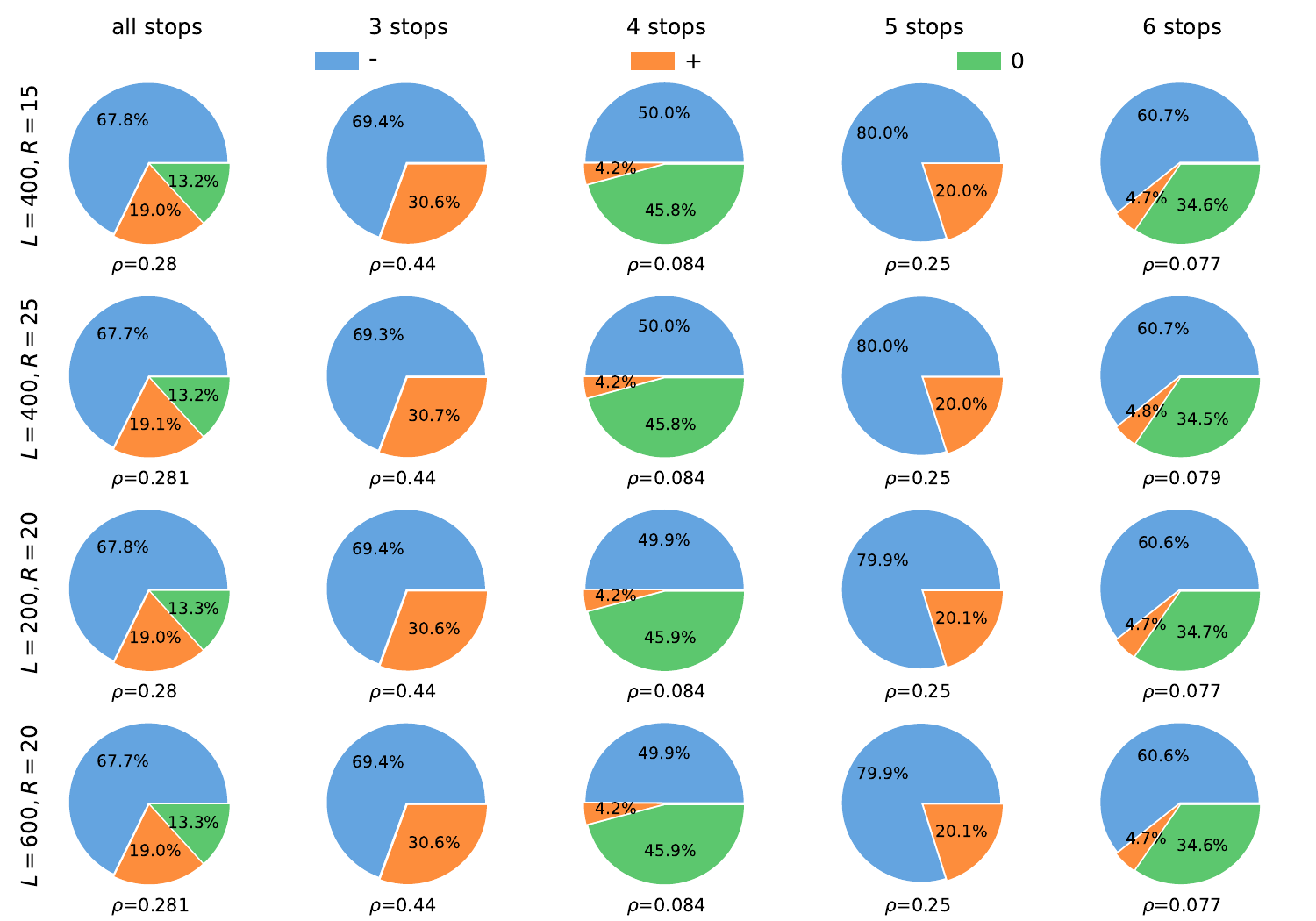}
\caption[The output of the d-mix model in different values of $R$ and $L$.]{{\textbf {The output of the d-mix model for different values of $R$ and $L$.}}
Fraction of positive $H(t) > 0$, negative $H(t) < 0$ and balanced $H(t) = 0$ trajectories with $3$, $4$, $5$, $6$ stops and for all the modeled trajectories for different values of $R$ and $L$.
}
\label{fig16}
\end{figure}

In the main text we performed numerical simulations for an idealized circular city of radius $R = 20 \, km$ in a square of side $L = 400 \, km$.
Here we repeat the simulations with different values of $R$ and $L$. As shown in Fig. \ref{fig16}, we see that the choice of $R$ and $L$ does not affect our results.

\section{The empirical mobility field }
\begin{figure}
\centering
\includegraphics[width=8cm]{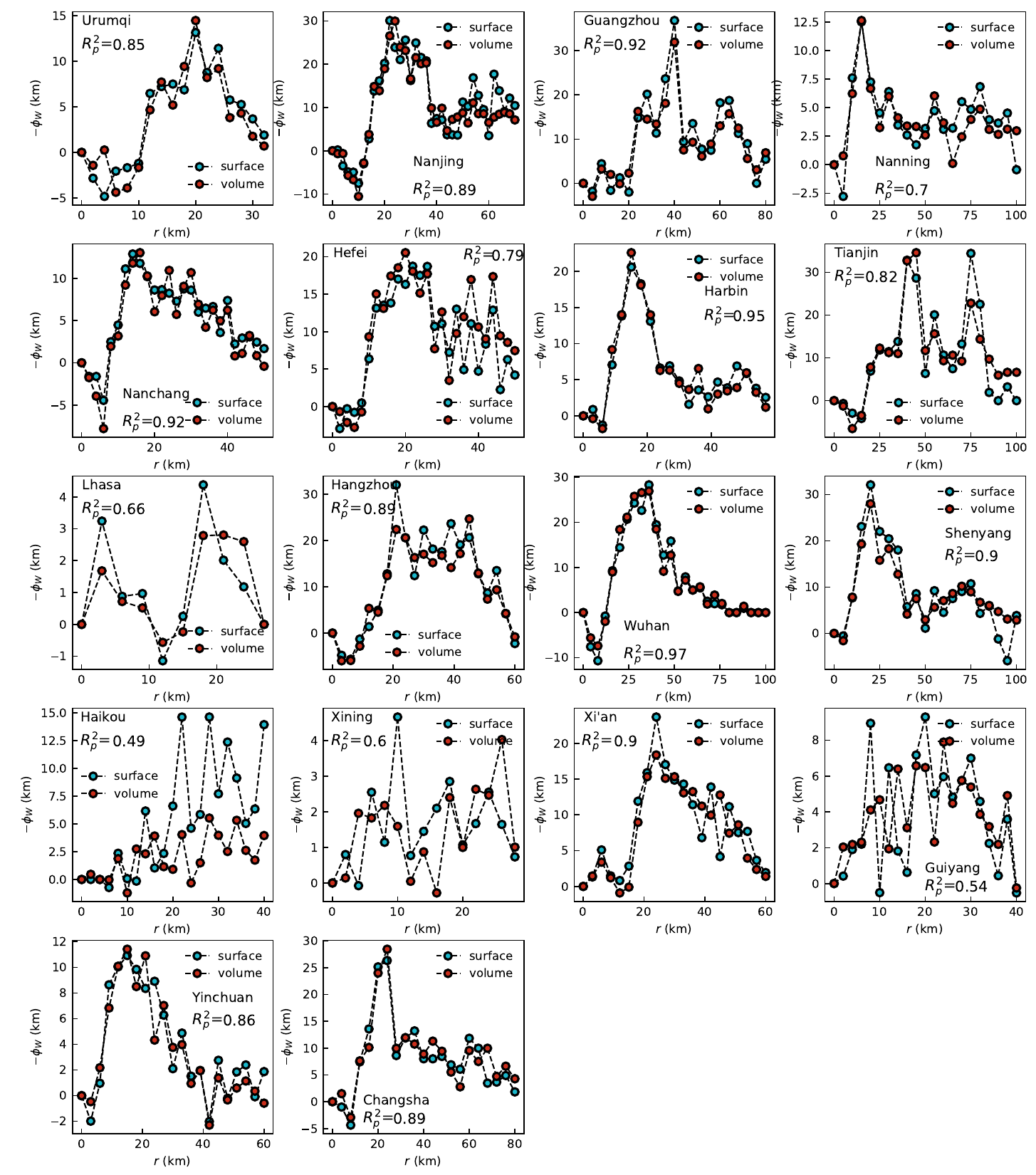}
\caption[Fluxes of the empirical field in Chinese cities]{{\textbf {Fluxes of the empirical field in Chinese cities.}} 
We consider trajectories in the remaining 18 Chinese cities not considered in the main text.
Fluxes of the empirical field $\vv{w}$ as a function of the distance to the city center $r$. The blue symbols correspond to the flux calculated as a surface integral and the red ones to the volume integral of the field divergence. The coefficient $R_p^2$ is obtained from a Pearson correlation of one flux with the other along the distance $r$.
}
\label{fig17}
\end{figure}

In main text Fig. 6, we see that the empirical fields generated in Beijing, Shanghai and Chengdu fulfil the Gauss' Theorem as well. Results for the other cities (18 cities) are included in Fig. \ref{fig17}. The centers of the cities for the flux calculations used to check  Gauss' Theorem have been taken at
locations shown in Table \ref{table:tables2}.
We can find that the empirical fields generated 18 cities all fulfil the Gauss' Theorem as well.  Note that  the value of coefficient $R_p^2$  in Haikou, Lhasa and Xining  we obtained is a little small, this because the number of trucks in  three cities is relatively rare, see Table \ref{table:tables2}.

\begin{figure}
\centering
\includegraphics[width=8cm]{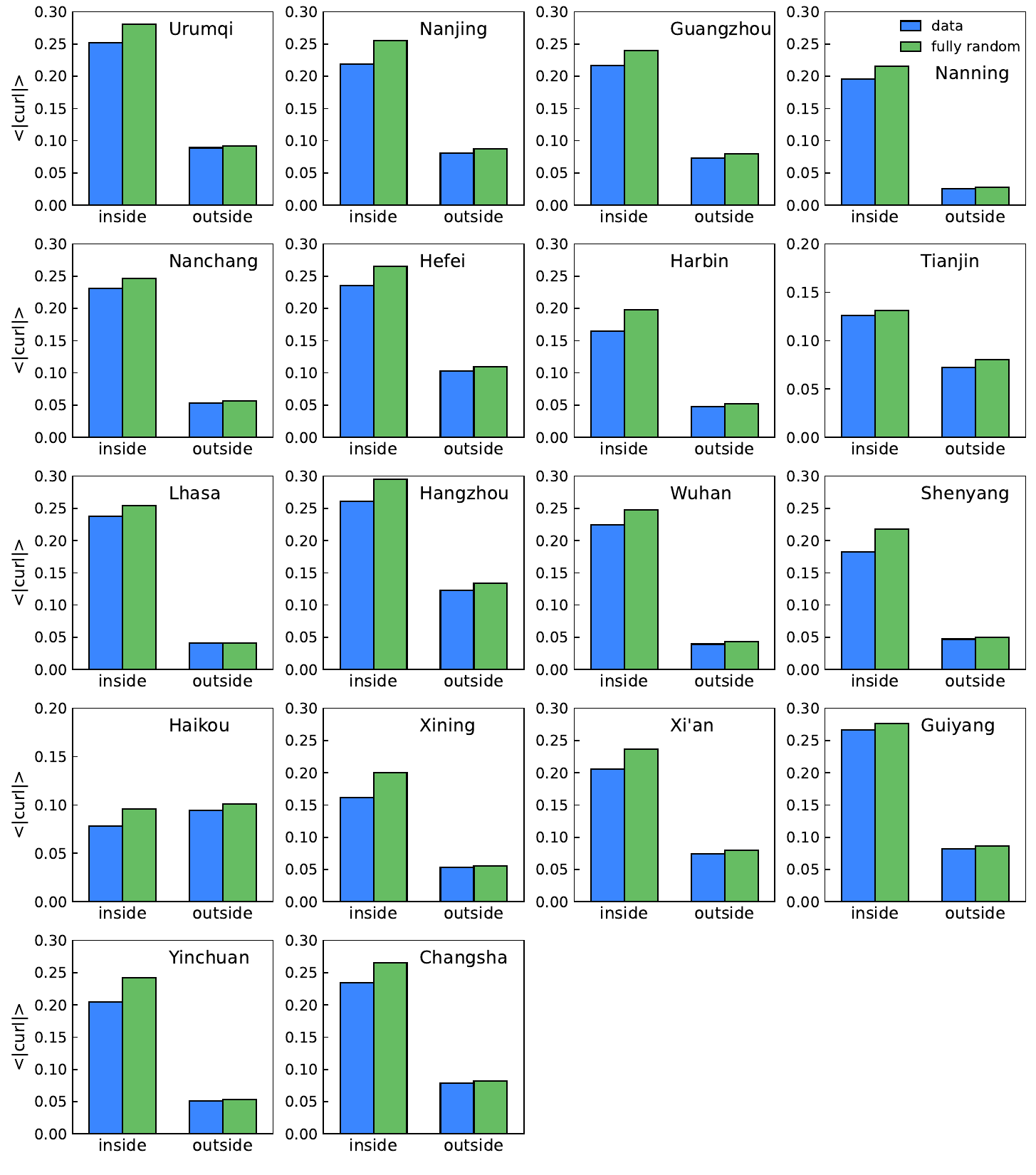}
\caption[Average
module of the curl in  Chinese cities]{{\textbf {Average
module of the curl in Chinese cities.}} 
We consider trajectories in the remaining 18 Chinese cities not considered in the main text.
Average module of the curl, comparing fully-random model and empirical data. The separation between in and out is defined by the radius $r_c$ at which $\Phi = \Phi_{\rm max}/2$. The inner part is enclosed by the radius $r_c$, the remaining area is the outer part, which ends at $r = 50 \, km$. After this distance, trip vectors become very sparse and statistics become non-representative.
}
\label{fig18}
\end{figure}

\begin{figure}
\centering
\includegraphics[width=8cm]{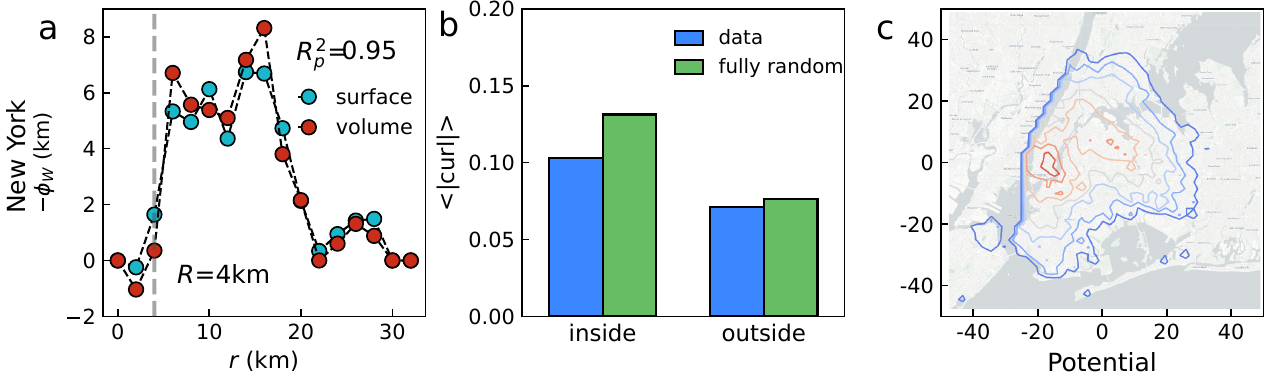}
\caption[ Empirical vector field in NYC.] {{\textbf {Empirical vector field in NYC.}} In {\bf a}, fluxes of the empirical field $\vv{w}$ as a function of the distance to the city center $r$. The blue symbols correspond to the flux calculated as a surface integral and the red ones to the volume integral of the field divergence. The coefficient $R_p^2$ is obtained from a Pearson correlation analysis of one flux with the other along distance $r$.  {\bf b} Average module of the curl as a function of the distance to the city center $r$. Inbox, comparing the empirical and randomized model average curl modules inside the city and outside. The separation between inside-outside has been arbitrarily establish at $R = 4 \, km$ in NYC.  {\bf c} Empirical equipotential contours for  NYC.}
\label{fig19}
\end{figure}

\begin{figure}
\centering
\includegraphics[width=8cm]{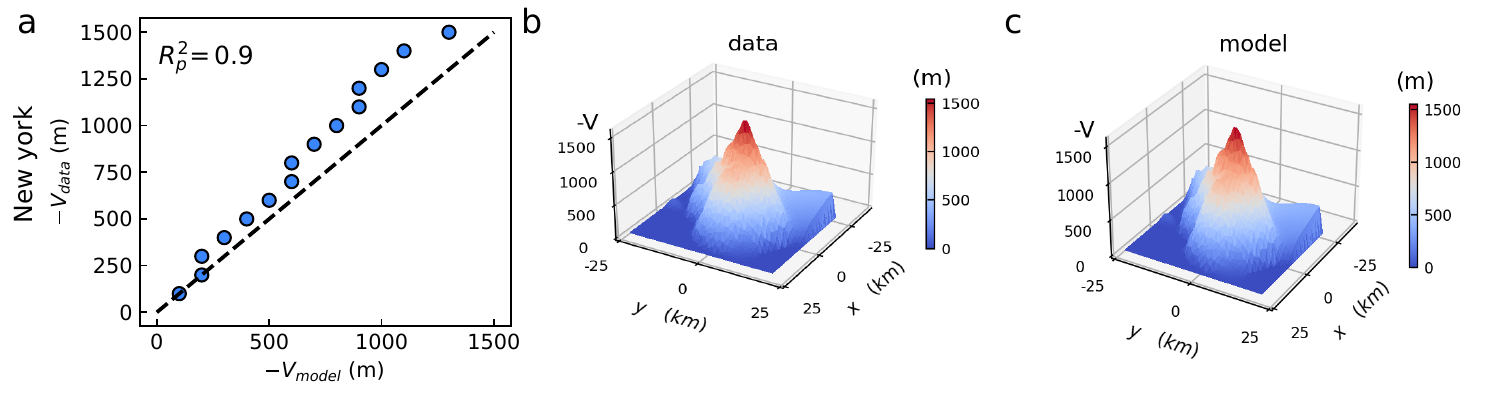}
\caption[Hybrid model predictions of empirical potentials]{{\textbf {Hybrid model predictions of empirical potentials} } {\bf a} Correlation plots between the hybrid d-mix model and the empirical potentials, yielding ${R_p}^2=0.9$. {\bf b} Empirical potential in the space and {\bf c} the hybrid d-mix model predictions.  }
\label{figs20}
\end{figure}

\section{Fit of the average distance $\ell_c$ for the d-mix model}
The d-mix model has a parameter $\ell_c$ that directly determines whether the agent takes the optimal or a random route. 
For a given empirical dataset, we rely on the track orientation metric $\rho$ and use the Eq. (11) in the main text to estimate $\ell_c$. 

\begin{figure}
\centering
\includegraphics[width=8cm]{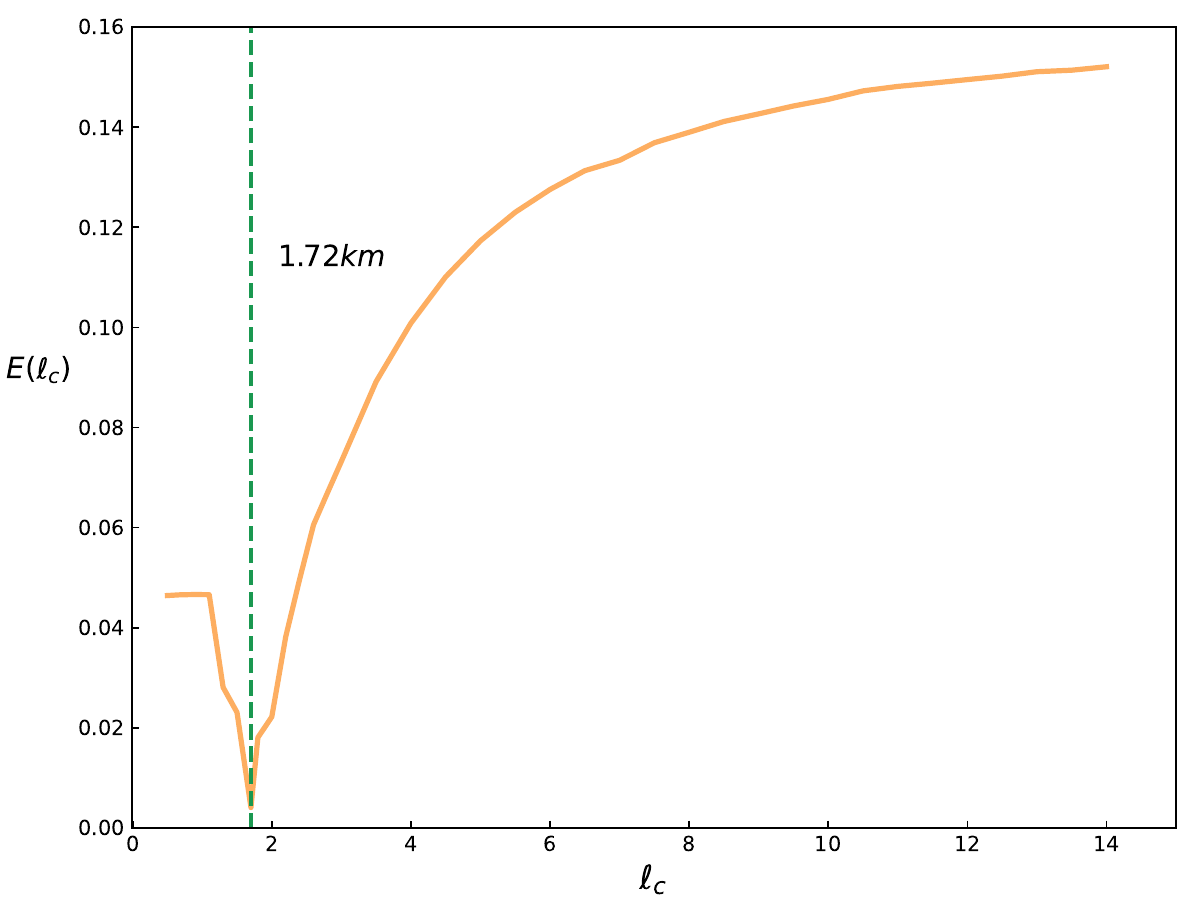}
\caption[Mean absolute error between modeled and observed ratio $\rho$ as a function of average distance $\ell_c$.]{{\textbf {Mean absolute error between modeled and observed ratio $\rho$ as a function of average distance $\ell_c$.} The x-axis represents the average distance $\ell_c$, and the y-axis represents the mean absolute error between the model and the empirical $\rho$. }  }
\label{fig21}
\end{figure}

In Fig. \ref{fig21}, we show that the mean absolute error $E(\ell_c)$ registers a minimum for an average distance $\ell_c = 1.72\ km$. For lower values of $\ell_c$, individuals optimize their trajectory even for very short trips. Vice versa, for higher values of $\ell_c$, individuals take random routes even for long trips.

\end{document}